\documentclass[12pt]{article}%
\usepackage[nosort]{cite}
\usepackage{graphicx}
\usepackage{multicol}
\usepackage{amsfonts}
\usepackage{amssymb}
\usepackage{amsmath}
\usepackage{heck}
\usepackage{afterpage}
\usepackage{setspace}
\usepackage{verbatim}
\usepackage{color}
\usepackage{longtable}
\usepackage{float}
\usepackage{subcaption}
\usepackage{epsfig}
\usepackage{epstopdf}
\usepackage{adjustbox}
\usepackage{tikz}
\usepackage[margin=1in]{geometry}
\usepackage{titletoc}
\usepackage{hyperref}%
\usepackage{mathrsfs}
\usepackage{cancel}

\providecommand{\U}[1]{\protect\rule{.1in}{.1in}}
\newsavebox{\mysavebox}

\hypersetup{colorlinks,citecolor=black,filecolor=black,linkcolor=black,urlcolor=black,pdftex}
\usetikzlibrary{decorations.markings}

\numberwithin{equation}{section}

\newcommand{\Tr}{\, {\rm Tr}}

\newcommand{\be}{\begin{equation}}
\newcommand{\ee}{\end{equation}}

\newcommand{\dd}{\mathrm{d}}

\usepackage{tikz}
\usetikzlibrary{arrows}
\usetikzlibrary{arrows.meta}
\usetikzlibrary{arrows,decorations.pathmorphing}
\usetikzlibrary{shapes.geometric,calc,arrows, positioning,shapes.misc,decorations.markings}
\tikzset{
  big arrow/.style={
    decoration={markings,mark=at position 1 with {\arrow[scale=2,#1]{>}}},
    postaction={decorate},
    shorten >=0.4pt},
  big arrow/.default=black}

\pgfdeclarelayer{edgelayer}
\pgfdeclarelayer{nodelayer}
\pgfsetlayers{edgelayer,nodelayer,main}
\tikzstyle{none}=[inner sep=0pt]

\tikzstyle{NodeCross}=[draw, shape=circle, cross out, inner sep=0pt, minimum size=6pt,line width=0.25mm]
\tikzstyle{Circle}=[draw, shape=circle, black, fill=black, inner sep=0pt, minimum size=6pt]
\tikzstyle{circle}=[draw, shape=circle, black, fill=black, inner sep=0pt, minimum size=16pt]
\tikzstyle{Star}=[draw, shape=star, fill=red, star points=8, inner sep=0pt, minimum size=8pt]
\tikzstyle{CircleRed}=[draw, shape=circle, black, fill=red, inner sep=0pt, minimum size=6pt]
\tikzstyle{StarP}=[draw={rgb,255: red,128; green,0; blue,128}, shape=star, fill={rgb,256: red,128; green,0; blue,128}, star points=8, inner sep=0pt, minimum size=12pt]
\tikzstyle{ShadedCircRed}=[draw=red, shape=circle, fill={rgb, 255: red,255; green,114; blue, 118}, inner sep=0pt, minimum size=80pt, line width=0.5mm, fill opacity=0.2]
\tikzstyle{ShadedCircRed2}=[draw=red, shape=circle, fill={rgb, 255: red,255; green,114; blue, 118}, inner sep=0pt, minimum size=10pt]
\tikzstyle{ShadedCircRed3}=[draw=black, shape=rectangle, fill={rgb, 255: red,255; green,114; blue, 118}, inner sep=0pt, minimum size=113pt, line width=0.25mm]
\tikzstyle{ShadedCirc}=[draw=red, shape=circle, fill=white, inner sep=0pt, minimum size=45pt,  fill opacity=1.0,  line width=0.5mm]
\tikzstyle{CircleBlue}=[draw, shape=circle, fill=blue, inner sep=0pt, minimum size=6pt]
\tikzstyle{BigCirclePurple}=[draw, shape=circle, fill={rgb,255: red,191; green,0; blue,191}, inner sep=0pt, minimum size=9pt]
\tikzstyle{CirclePurple}=[draw, shape=circle, fill={rgb,255: red,191; green,0; blue,191}, inner sep=0pt, minimum size=8pt]
\tikzstyle{EmptyCircle}=[draw, shape=circle, inner sep=0pt, minimum size=4pt]
\tikzstyle{GreenCircle}=[draw, shape=circle,  fill={rgb,255: red,80; green,200; blue,120}, inner sep=0pt, minimum size=8pt]
\tikzstyle{BrownCircle}=[draw, shape=circle,  fill={rgb,255: red,210; green,105; blue,30}, inner sep=0pt, minimum size=8pt]
\tikzstyle{CirclePurpleSmall}=[draw, shape=circle, fill={rgb,255: red,191; green,0; blue,191}, inner sep=0pt, minimum size=4pt]
\tikzstyle{BigCircleGreen}=[draw, shape=circle, fill={rgb,255: red,0; green,191; blue,0}, inner sep=0pt, minimum size=12pt]
\tikzstyle{BigCircleBlue}=[draw, shape=circle, fill={rgb,255: red,0; green,0; blue,191}, inner sep=0pt, minimum size=12pt]
\tikzstyle{BigCircleRed}=[draw, shape=circle, fill={rgb,255: red,191; green,0; blue,0}, inner sep=0pt, minimum size=12pt]
\tikzstyle{BrownCircleSmall}=[draw, shape=circle,  fill={rgb,255: red,210; green,105; blue,30}, inner sep=0pt, minimum size=6pt]
\tikzstyle{SmallCirclePurple}=[draw, shape=circle, fill={rgb,255: red,191; green,0; blue,191}, inner sep=0pt, minimum size=4pt]
\tikzstyle{SmallCircleRed}=[draw, shape=circle, fill={rgb,255: red,191; green,0; blue,0}, inner sep=0pt, minimum size=4pt]
\tikzstyle{SmallCircleGreen}=[draw, shape=circle, fill={rgb,255: red,0; green,191; blue,0}, inner sep=0pt, minimum size=4pt]

\tikzstyle{DashedLine}=[-, densely dashed, line width=0.25mm]
\tikzstyle{DottedLine}=[-, dotted, line width=0.25mm]
\tikzstyle{ThickLine}=[-, line width=0.25mm]
\tikzstyle{ArrowLineRight}=[-, -{Stealth[scale=1.25]}, line width=0.25mm, scale=5]
\tikzstyle{ArrowLineRed}=[-, draw={rgb,255: red,191; green,0; blue,0}, -{Stealth[scale=1.75]}, line width=0.1mm, scale=5]
\tikzstyle{RedLine}=[-, draw={rgb,255: red,191; green,0; blue,0}, fill=none, line width=0.5mm]
\tikzstyle{DashedLineThin}=[-, densely dashed, line width=0.125mm, fill=none, draw=black]
\tikzstyle{DottedRed}=[-, dotted, draw={rgb,255: red,191; green,0; blue,0}, fill=none, line width=0.25mm]
\tikzstyle{DashedRed}=[-, densely dashed, draw={rgb,255: red,191; green,0; blue,0}, fill=none, line width=0.25mm]
\tikzstyle{BlueLine}=[-, draw={rgb,255: red,0; green,0; blue,191}, fill=none, line width=0.5mm]
\tikzstyle{DottedBlueLine}=[-, dotted,draw={rgb,255: red,0; green,0; blue,191}, fill=none, line width=0.5mm]
\tikzstyle{ArrowLineBlue}=[-, draw={rgb,255: red,0; green,0; blue,191}, -{Stealth[scale=1.75]}, line width=0.1mm, scale=5]
\tikzstyle{GreenDoubleArrow}=[<->, draw={rgb,155: red,0; green,255; blue,0},  line width= 0.5mm, scale=5]
\tikzstyle{RedDoubleArrow}=[<->, draw={rgb,255: red,255; green,0; blue,0},  line width= 0.5mm, scale=5]
\tikzstyle{BlueDottedLight}=[-, dotted, draw={rgb,255: red,0; green,0; blue,191}, fill=none, line width=0.3mm]
\tikzstyle{BrownLine}=[-, draw={rgb,255: red,210; green,105; blue,30}, fill=none, line width=0.5mm]
\tikzstyle{DottedRed}=[-, dotted, draw={rgb,255: red,191; green,0; blue,0}, fill=none, dotted, line width=0.5mm]
\tikzstyle{DottedPurple}=[-, dotted, draw={rgb,255: red,191; green,0; blue,191}, fill=none, dotted, line width=0.5mm]
\tikzstyle{BlueDottedLight}=[-, dotted, draw={rgb,255: red,0; green,0; blue,191}, fill=none, line width=0.5mm]
\tikzstyle{ArrowLinePurple}=[-, draw={rgb,255: red,191; green,0; blue,191}, -{Stealth[scale=1.75]}, line width=0.5mm, scale=5]
\tikzstyle{DashedLineGreen}=[-, densely dashed, draw={rgb,255: red,74; green,103; blue,65}, line width=0.25mm]
\tikzstyle{LineGreen}=[-, draw={rgb,255: red, 74; green,200; blue,65}, line width=0.5mm]
\tikzstyle{ArrowLineGreen}=[-, draw={rgb,255: red,0; green,191; blue,0}, -{Stealth[scale=1.75]}, line width=0.5mm, scale=5]
\tikzstyle{GreenLine}=[-, draw={rgb,255: red,0; green,191; blue,0}, fill=none, line width=0.5mm]
\tikzstyle{PurpleLine}=[-, draw={rgb,255: red,191; green,0; blue,191}, fill=none, line width=0.5mm]
\tikzstyle{PPurpleLine}=[-, draw={rgb,255: red,191; green,0; blue,191}, fill=none, line width=2.5mm]
\tikzstyle{DPurpleLine}=[-, dotted, draw={rgb,255: red,191; green,0; blue,191}, fill=none, line width=0.5mm]
\tikzstyle{SBrownLine}=[-, draw={rgb,255: red,191; green,0; blue,191}, fill=none, opacity=0.35, line width=2.5mm]

\tikzset{snake it/.style={decorate, decoration=snake}}

\usetikzlibrary{decorations.markings}
\usetikzlibrary{hobby}

\tikzset{
dashstar/.style={
 dash pattern=on 5pt off 5pt,
 postaction={
  decorate,
  decoration={
   markings,
   mark=between positions 9pt and 1 step 10pt with {
     \node[color=red] {*};
   }
  }
 }
},
dashstarstar/.style={ % from marmot's comments
 dash pattern=on 5pt off 10pt,
 postaction={
   decorate,
   decoration={
     markings,
     mark=between positions 10pt and 1
          step 15pt
           with {
            \node at (-2pt,0pt) {\pgfuseplotmark{star}};
            \node at (2pt,0pt) {\pgfuseplotmark{star}};
           }
   }
 }
}
}
\usepackage{pgfplots}
\pgfplotsset{compat=1.16}

\newcommand{\ba}{\begin{aligned}}
\newcommand{\ea}{\end{aligned}}

\begin{document}

\begin{flushright}
    UUITP-05/25
\end{flushright}

\date{January 2025}

\title{Metric Isometries, Holography, and \\[4mm] Continuous Symmetry Operators}

\institution{PENN}{\centerline{$^{1}$Department of Physics and Astronomy, University of Pennsylvania, Philadelphia, PA 19104, USA}}
\institution{PENNmath}{\centerline{$^{2}$Department of Mathematics, University of Pennsylvania, Philadelphia, PA 19104, USA}}
\institution{Maribor}{\centerline{$^{3}$Center for Applied Mathematics and Theoretical Physics, University of Maribor, Maribor, Slovenia}}
\institution{Uppsala}{\centerline{$^{4}$Department of Physics and Astronomy, Uppsala University, Box 516, SE-75120 Uppsala, Sweden}}

\authors{
Mirjam Cveti\v{c}\worksat{\PENN,\PENNmath,\Maribor}\footnote{e-mail: \texttt{cvetic@physics.upenn.edu}},
Jonathan J. Heckman\worksat{\PENN,\PENNmath}\footnote{e-mail: \texttt{jheckman@sas.upenn.edu}},\\[4mm]
Max H\"ubner\worksat{\Uppsala}\footnote{e-mail: \texttt{max-elliot.huebner@physics.uu.se}}, and
Chitraang Murdia\worksat{\PENN}\footnote{e-mail: \texttt{murdia@sas.upenn.edu}}
}

\abstract{In the AdS/CFT correspondence, a topological symmetry operator of the boundary
CFT is dual to a dynamical brane in the gravitational bulk. Said differently, this predicts a
dynamical brane for every global symmetry of the boundary CFT.
We analyze this correspondence for continuous symmetries which arise from a
consistent truncation of isometries on the ``internal'' factor $X$ of
$\mathrm{AdS} \times X$.  In the extra-dimensional geometry,
these branes are associated with various metric singularities and do not arise from wrapped D-branes.
Boosts relate configurations interpreted as topological symmetry operators and heavy defects in the CFT.
From the perspective of the $\mathrm{AdS}$ factor,
with gravity and bulk gauge fields, these are codimension two Gukov-Witten-like vortex configurations which are the
gravity duals of 0-form symmetry operators. These effective branes come with an asymptotic
tension and size which is also fully fixed by bulk dynamics. We use this higher-dimensional
perspective to determine properties of the worldvolume theory for these branes. We also discuss how these considerations
generalize to more general QFTs engineered via string theory which need not possess a semi-classical gravity dual.
}

\maketitle

\enlargethispage{\baselineskip}

\setcounter{tocdepth}{2}

\tableofcontents

\newpage

\section{Introduction}

One of the core features of the AdS/CFT correspondence is that gauge symmetries of the bulk correspond to global symmetries of the boundary theory.
In this context, it is natural to determine the bulk dual of a topological symmetry operator, as associated with a generalized symmetry \cite{Gaiotto:2014kfa}.
Recent evidence from top-down approaches to quantum gravity and holography has established that in many cases of interest, this bulk dual is a dynamical brane.
Indeed, starting from a dynamical brane, the passage to the boundary freezes out its dynamical sector, leaving behind only a topological field theory associated with a given symmetry operator \cite{Apruzzi:2022rei, GarciaEtxebarria:2022vzq, Heckman:2022muc}.
A bottom-up approach based on bulk reconstruction and subregion-subregion duality was recently used to provide a novel proof of this statement based on quite minimal assumptions \cite{Heckman:2024oot}.
As a corollary, this also implies that there are no global symmetries in any holographic spacetime with subregion-subregion duality.\footnote{This is complementary (and extends) the considerations presented in \cite{Harlow:2018jwu, Harlow:2018tng}. For related recent discussions on the absence of global symmetries in gravity, see \cite{Heckman:2024oot, Bah:2024ucp}.}

One consequence of these considerations is that \textit{any} symmetry operator in a boundary CFT$_{D}$ predicts the existence of a bulk dual brane.
Some of these are recognizable as wrapped D-branes, but more generally there is no need for this to be the case.
For example, the bulk dual of a charge conjugation symmetry operator is a more general object \cite{Dierigl:2023jdp}.\footnote{There is a clear connection to the Swampland cobordism conjecture \cite{McNamara:2019rup} which argues based on the absence of global
symmetries / spectrum completeness that dynamical branes must often be added to a gravitational system.
See also \cite{Dierigl:2020lai,Debray:2021vob, Blumenhagen:2021nmi, Buratti:2021fiv, Dierigl:2022reg,
Blumenhagen:2022bvh, Debray:2023yrs, Dierigl:2023jdp}.}

A case of particular interest is that of continuous 0-form symmetries for a semi-simple Lie group $G$.
In the case of stringy realizations of the AdS$_{D+1}$/CFT$_{D}$ correspondence, the gravitational dual is of the form AdS$_{D+1}\times X$, and the isometry group of $X$ determines (after a suitable lift to include the action on spinors) a bulk gauge group $G$.
In the boundary theory, this is characterized by a 0-form global symmetry.
For example, the R-symmetry of a superconformal field theory (SCFT)\ with a gravity dual is a subgroup of the isometry group of $X$.
In this regard, a natural question is to determine the gravity dual to the corresponding topological symmetry operators.

Our aim in this paper will be to explicitly identify the symmetry operators and defects associated with such isometries.
We study this issue in the context of the AdS/CFT correspondence as well as in the context of the expected symmetry topological field theory (SymTFT) / symmetry theory (SymTh) governing such continuous symmetries. Our considerations also apply more broadly to QFTs (holographic or not) engineered via string theory where isometries of the local geometry $\mathrm{Cone}(X)$ also enact symmetries of the QFT localized at the tip of the cone.

Recall that in the SymTFT\,/\,SymTh formalism,\footnote{See, e.g., \cite{Reshetikhin:1991tc, Turaev:1992hq, Barrett:1993ab, Witten:1998wy, Fuchs:2002cm, Kirillov:2010nh, Kapustin:2010if, Kitaev:2011dxc, Fuchs:2012dt, Freed:2012bs, Heckman:2017uxe, Freed:2018cec, Gaiotto:2020iye, Apruzzi:2021nmk, Freed:2022qnc, Kaidi:2022cpf, Brennan:2024fgj, Heckman:2024oot, Antinucci:2024zjp, Bonetti:2024cjk, Apruzzi:2024htg, Braeger:2024jcj, Heckman:2024zdo, Cvetic:2024dzu, GarciaEtxebarria:2024jfv, Najjar:2024vmm}).}
one specifies a $D$-dimensional QFT$_{D}$ and then ``decompresses'' it
to a $(D+1)$-dimensional system with two boundaries, namely one boundary where we have a relative QFT (in the sense of \cite{Freed:2012bs, Freed:2022qnc}) and a topological boundary condition where one specifies the global form of the absolute QFT.

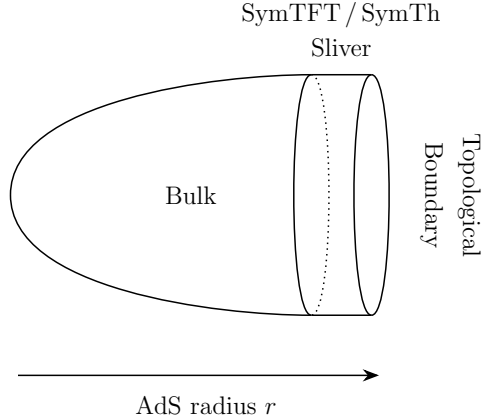
\begin{figure}
\centering
\scalebox{0.8}{
\begin{tikzpicture}
	\begin{pgfonlayer}{nodelayer}
		\node [style=none] (0) at (3, 2) {};
		\node [style=none] (1) at (3, -2) {};
		\node [style=none] (2) at (2, -2) {};
		\node [style=none] (3) at (2, 2) {};
		\node [style=none] (4) at (-3, 0) {};
		\node [style=none] (5) at (0, 0) {Bulk};
		\node [style=none] (6) at (4, 0) {\rotatebox{-90}{Boundary}};
		\node [style=none] (6) at (4.625, 0) {\rotatebox{-90}{Topological}};
		\node [style=none] (7) at (2.5, 3) {SymTFT\,/\,SymTh};
		\node [style=none] (7) at (2.5, 2.45) {Sliver};
		\node [style=none] (8) at (3.125, -3) {};
		\node [style=none] (9) at (-2.875, -3) {};
		\node [style=none] (10) at (0.25, -3.5) {AdS radius $r$};
		\node [style=none] (11) at (0.25, -4) {};
	\end{pgfonlayer}
	\begin{pgfonlayer}{edgelayer}
		\draw [style=ThickLine, in=180, out=-180, looseness=0.25] (0.center) to (1.center);
		\draw [style=ThickLine, in=-180, out=-180, looseness=0.25] (3.center) to (2.center);
		\draw [style=ThickLine, in=0, out=0, looseness=0.25] (0.center) to (1.center);
		\draw [style=DottedLine, in=0, out=0, looseness=0.25] (3.center) to (2.center);
		\draw [style=ThickLine] (3.center) to (0.center);
		\draw [style=ThickLine] (2.center) to (1.center);
		\draw [style=ThickLine, in=-90, out=180, looseness=0.75] (2.center) to (4.center);
		\draw [style=ThickLine, in=-180, out=90, looseness=0.75] (4.center) to (3.center);
		\draw [style=ArrowLineRight] (9.center) to (8.center);
	\end{pgfonlayer}
\end{tikzpicture}}
\caption{The physical boundary condition to the symmetry theory is the complete AdS bulk, which is equivalent to the dual relative CFT. The symmetry theory lives in an asymptotic sliver, bounded on the other side by the topological boundary condition.}
\label{fig:sliver}
\end{figure}

We present a general procedure for building symmetry operators in the bulk starting from non-topological heavy defects of the gravity dual. Key examples of such defects were recently derived in \cite{Arav:2024exg, Bomans:2024vii}. Much as in \cite{Heckman:2024oot} (which builds on the perspective developed in \cite{Aharony:1998qu, Witten:1998wy, Maldacena:2001ss}) we view the SymTh$_{D+1}$ as a small sliver of the full gravitational bulk living in AdS$_{D+1}$ (see figure \ref{fig:sliver}).\footnote{Much as in \cite{Heckman:2024oot}, the appearance of the sliver is in line with the ``extrapolate'' dictionary of AdS/CFT \cite{Susskind:1998dq, Banks:1998dd} in which one pushes operators to the boundary, as opposed to the ``differentiate'' dictionary of \cite{Gubser:1998bc, Witten:1998qj} in which one differentiates a generating function with respect to a source. In the AdS/CFT correspondence, the two dictionaries are expected to be equivalent (see \cite{Harlow:2011ke} for a discussion along these lines) and so we will not belabor this point further.} From this perspective, one can first build a heavy defect in AdS and then boost it using the isometries of AdS$_{D+1}$ so that its radial profile is concentrated close to the conformal boundary.
Detaching this object from the boundary can then in principle be accomplished at the expense of introducing (possibly trivial) operators that stretch back to the conformal boundary.
%The boundary of the flux tube defines an object which becomes the corresponding symmetry operator in the dual CFT${_D}$.

Our main focus in the present work will be symmetry and defect operators associated with a continuous 0-form symmetry in the boundary CFT$_D$.
In this case, the symmetry operator is a topological operator of codimension-1 in the boundary theory. In the bulk, this becomes a codimension-2 object (which can sometimes attach back to the boundary) \cite{Heckman:2024oot}. To construct these symmetry operators and their bulk duals, we start with a bulk defect which extends radially in the AdS$_{D+1}$ spacetime, terminating on the conformal boundary. Applying a suitable boost pushes this to the SymTFT sliver, and when paired with an additional defect can be used to fully detach it from the boundary. We use this construction technique to read off local data intrinsic to these bulk dual objects, including its tension as well as its worldvolume topological sector via the SymTFT$_{\mathrm{defect}}$ for the defect.

We shall be interested in constructing the symmetry operators associated with the isometries of $X$ in the gravity dual AdS$_{D+1} \times X$.
For a choice of $g \in G = \mathrm{Isom}(X)$ there is a corresponding Killing vector $\xi_{g}$. We focus attention on the case of continuous isometries which are connected to the identity. As such, we also know that $g = \exp(i \alpha)$, with $\alpha \in \mathfrak{g}$ the Lie algebra of $G$. Observe that for a generic $\lambda \in [0,1)$ the element $g_{\lambda} = \exp(i \lambda \alpha)$ determines a family of $S^1$'s.\footnote{This is immediate when $g$ is of infinite order. When $g$ is of finite order observe that one still has a distinguished set of circles.} For ease of exposition, we primarily focus on the symmetries associated with a $U(1) \subset G$, but we comment that the method of construction works equally well when $G$ is abelian or non-abelian.

Focussing then on the case of $G = U(1)$, observe that Kaluza-Klein (KK) momenta along these distinguished $S^1$'s build up heavy particles in the corresponding AdS$_{D+1}$; these can be interpreted as electric line operators associated with the corresponding KK gauge field. In the SymTFT for this continuous symmetry one expects to find magnetic dual objects which topologically link with these lines in the bulk. These are associated with codimension-2 defects in the AdS$_{D+1}$ bulk which descend to the codimension-1 symmetry defects of the boundary CFT$_D$.

There is a natural candidate for this magnetic dual object; it is the KK monopole configuration associated with a given $S^1$ foliation of $X$. Physically, this is realized in the bulk via a flux of the form:
\begin{equation}\label{eq:bulkF2}
F_{2} = 2 \pi \alpha \delta^{(2)}(\Sigma_{D-1}),
\end{equation}
namely a flux configuration which is localized on the codimension-2 subspace $\Sigma_{D-1}$.
These flux configurations end on the CFT$_D$ and as such need not be properly quantized.

This sort of field configuration also arises naturally in the CFT$_D$. Given a global symmetry $G$, we can introduce a background gauge field $a$ with field strength:
\begin{equation}
f_{2} = 2 \pi \alpha \delta^{(2)}(\Sigma_{D-2}),
\end{equation}
where in connecting with the discussion of line (\ref{eq:bulkF2}) we simply view $\Sigma_{D-1}$ as terminating on $\Sigma_{D-2}$. Indeed, these sorts of delta-function localized contributions arise from singular gauge field configurations such as $a = \frac{1}{i} h^{-1} \dd h$ with $h = \exp(i \alpha \phi)$, in the obvious notation. This specifies a Gukov-Witten-like codimension-2 defect for the flavor symmetry which extends in the radial direction of the bulk AdS$_{D+1}$.

Geometrically, the choice of background KK gauge field $A_{\mathrm{KK}}$ in the AdS$_{D+1}$ spacetime lifts in AdS$_{D+1} \times X$ to a choice of metric data in which the topological space $X$ now fibers non-trivially over AdS$_{D+1}$. Thankfully, the relevant field configurations have already been constructed in many cases of interest in \cite{Arav:2024exg, Bomans:2024vii}. In particular, properties of the resulting defect such as the tension / conical deficit angle are specified purely by the choice of $g \in G$.

From such solutions, the general boosting procedure can be directly used to build a class of symmetry operators.
In this setting, all of the radial dependence of the original symmetry operator is now pushed into a direction that runs parallel to a direction filled by the CFT$_{D}$. The appearance of a non-trivial radial dependence is interpreted, in the boosted brane, as non-topological contributions to the worldvolume action. In particular, in the limit where we push this dynamical brane into the conformal boundary, all of this non-topological data is stripped off, realizing the symmetry operator in question.

Now, precisely because the defect in question is realized via a singular / localized geometry, we can treat it as engineering a QFT in its own right. In particular, to figure out data such as the topological subsector associated with generalized Wess-Zumino terms (in analogy with the D-brane case) we also develop a general prescription for reading this data off directly from the accompanying SymTFT$_{\mathrm{defect}}$ for the defect brane.\footnote{We emphasize that this is the SymTFT$_{\mathrm{defect}}$ for the defect and \textit{not} the SymTFT$_{D+1}$ for the CFT$_D$.} This method is of broader interest since it allows us to extract topological couplings for a QFT directly from its associated SymTFT.

The considerations presented here generalize in a number of natural ways. While we primarily illustrate our considerations in the case of $\mathrm{AdS}_5 / \mathrm{CFT}_4$ pairs, it is clear that these considerations extend to the broader setting of $\mathrm{AdS}_{D+1} / \mathrm{CFT}_{D}$ pairs. Additionally, we can also dispense with the requirement that the string background realizing a $\mathrm{QFT}_{D}$ even
has a semi-classical gravity dual. Indeed, so long as we have a local background of the form $\mathbb{R}^{D-1,1} \times \mathrm{Cone}(X)$, then we expect the isometries of $X$ to yield non-trivial global symmetries in the localized QFT. To illustrate this point we leverage this geometric point of view to first engineer $\mathcal{N} = 4$ Super Yang-Mills theory directly in geometry, and then to use isometries of this background to realize duality / triality defects as special values of the complexified coupling $\tau$. This allows us to establish further properties of these defects, including their interplay with surface operators of the gauge theory.

The rest of this paper is organized as follows.
We begin in section \ref{sec:BOOST} by discussing in broad terms a general strategy for constructing symmetry operators
starting from radially extended defects. After this, in section \ref{sec:DEFECTS} we construct the defects and symmetry operators
associated with metric isometries of $X$ in backgrounds of the form AdS$_{D+1} \times X$. We also use this perspective to extract properties
such as the tension of the bulk duals to these symmetry operators. In section \ref{sec:PROPERTIES} we give a general prescription for reading off topological couplings directly from the SymTFT$_{\mathrm{defect}}$ from these defects. In section \ref{sec:NONHOLO} we show how these considerations naturally extend to examples without a holographic dual. We summarize and discuss some potential future directions of investigation in section \ref{sec:CONC}. In Appendix \ref{app:MASSIVE} we discuss the case of a broken symmetry associated with a massive bulk gauge field in AdS, where the putative symmetry operator (for the broken symmetry) is realized via a wrapped brane.

\section{Symmetry Operators via Boosted Defects} \label{sec:BOOST}

One of the important features of the SymTFT\,/\,SymTh formalism is that heavy defects and symmetry operators are on a similar footing from the perspective of the higher-dimensional bulk system.
This agrees well with expectations from holography where one expects all such objects to be associated with dynamical (i.e., fluctuating) branes.
Indeed, in the context of the SymTFT formalism, one specifies a topological boundary $\left\vert \text{top}\right\rangle $ and a physical boundary $\left\vert \text{phys}\right\rangle $.
The choice of topological boundary condition then dictates which bulk objects are heavy defects, and which are instead symmetry operators.
The main point is that the heavy defects stretch from the physical boundary to the topological boundary, whilst the topological operators do not attach to the
boundaries.\footnote{There are circumstances where the bulk symmetry operator cannot fully detach from the physical boundary.
For example, the symmetry generators of a non-abelian group typically do not detach.
The ones which can detach are labeled by a conjugacy class. See \cite{Cordova:2022rer, Heckman:2024oot} for further discussion on this point.\newline \indent  ~Another obstruction to detaching occurs when operators are twist/monodromy-operators in the sense of \cite{Kaidi:2022cpf, Arav:2024exg}. In this case, a ``branch cut", associated with the twist/monodromy, supporting topological terms can emanate from the defect and must be terminated elsewhere. See \cite{Heckman:2022xgu, Dierigl:2023jdp} where this occurs in constructions of defects utilizing 7-branes.}

There is a natural extension of these considerations to CFTs with a semi-classical gravity dual.
In that context, the physical boundary is better viewed as enlarging to AdS$_{D+1}$. See figure \ref{fig:sliver}.
Heavy defects then extend along the radial direction of AdS$_{D+1}$, while the branes associated with topological operators can be quasi-localized at a fixed radial position.

Starting from a heavy defect, we now explain how to use the isometries of AdS$_{D+1}$ to push it fully into the small topological sliver associated with the SymTFT$_{D+1}$. We can always detach this defect from the boundary at the expense of introducing a defect-anti-defect fusion product connecting back to the boundary.\footnote{In the case of abelian invertible symmetries, the defect-anti-defect fusion product is trivial.
See \cite{Heckman:2024oot} for further details on this point.}
%The end of this flux tube is the bulk dual of the symmetry operator.
The top-degree charge of the latter vanishes, and in this sense, the operator has been detached from the boundary. Often, the heavy defect will deform the semi-classical AdS$_{D+1}$ background, breaking the initial isometry group to a subgroup. In such cases, which are generic for high codimension defects, it is the broken isometries which relate heavy defect and symmetry operator configurations. These deformations are case-dependent, and we will idealize defects here as probes without back reaction, deferring the more careful treatment to section \ref{sec:Bulk} where they are analyzed using various consistent truncations.

We now show how to push a heavy defect close to the boundary. We first carry out the procedure
in Euclidean AdS$_{D+1}$ (i.e., on a topological ball) and then explain how this procedure works in Lorentzian AdS$_{D+1}$.
We first focus on the case $D > 2$ and then turn to the special case of AdS$_3$ and AdS$_2$ backgrounds since some of the details are
different in this case.

\subsection{Euclidean AdS$_{D+1}$ Boosting}

We start with a radially extended defect of Euclidean AdS$_{D+1}$ for $D > 2$ and show how to push it fully into the SymTFT sliver. This can be used to build a class of symmetry operators for the boundary CFT$_D$.

To this end, consider the embedding space for Euclidean AdS$_{D+1}$ in $\mathbb{R}^{1,D+1}$, as specified by the hypersurface:
\begin{equation}
    -L^{2}
    = -T^{2} + \left(  X^{1} \right)^{2} + \dots +\left( X^{D+1} \right)^{2} \, .
\end{equation}
Global coordinates that satisfy this constraint equation are given by
\begin{equation}
    T=L\cosh\rho
    \, ,
    \quad
    X^{j}=L\Omega^{j}\sinh\rho \, ,
    \qquad
    \text{with }
    \,
    \underset{j=1}{\overset{D+1}{\sum}} \, \Omega^{j}\Omega^{j}=1 \, ,
\end{equation}
in the usual notation.
Here, the constrained $\Omega^{j}$ parameterize an $S^{D}$, and the radial coordinate $\rho \in [0,\infty)$ ranges from the deep interior ($\rho \to 0$) to the conformal boundary ($\rho \to \infty$).
The argument we present can be adapted to more general boundary manifolds, so for now we leave it as $S^{D}$.

Suppose that we have a heavy defect that sweeps out a codimension-$p$ subspace AdS$_{d+1}$, where $d=D-p+1$.
Without loss of generality, we identify this AdS$_{d+1}$ subspace by setting $X^{1}=X^{2}=...=X^{p}=0$.\footnote{Note that the
codimension cannot be too high i.e., we require $p\leq D$,
otherwise the defect cannot extend along the radial direction.}
We push this defect away from the deep interior by performing a boost in the $(T,X^{p})$ plane.
In terms of the boosted coordinates we have:
\begin{equation}
    \begin{bmatrix}
    T_{\text{new}}\\
    X_{\text{new}}^{p}
    \end{bmatrix}
    =
    \begin{bmatrix}
    \cosh\beta & \sinh\beta\\
    \sinh\beta & \cosh\beta
    \end{bmatrix}
    \begin{bmatrix}
    T\\
    X^{p}%
    \end{bmatrix}
    \, .
\end{equation}
In particular, since $X^{p}=0$ along the defect, we learn that the defect is
now, after boosting, localized along the line:
\begin{equation}
    X_{\text{new}}^{p} = T_{\text{new}}\tanh\beta \, ,
\end{equation}
which in terms of the new global coordinates for Euclidean AdS$_{D+1}$ is:
\begin{equation}
    \Omega^{p} \tanh\rho = \tanh\beta \, .
\end{equation}
On the other hand, since $\left\vert \Omega^{p}\right\vert \leq1$, we learn
that $\left\vert \tanh\rho\right\vert \geq\tanh\beta$, with saturation of the
inequality only when $\left\vert \Omega^{p}\right\vert =1$. In particular,
this means that the defect has now been pushed to $\rho\geq\left\vert
\beta\right\vert $.

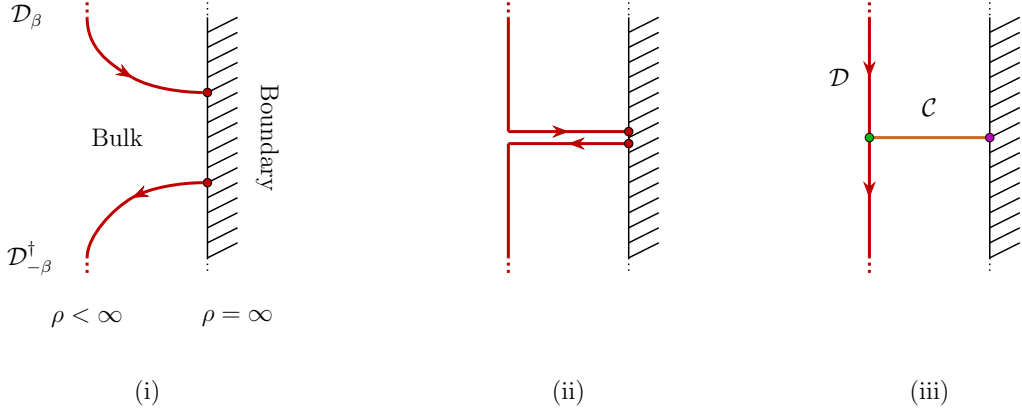
\begin{figure}
\centering
\scalebox{0.8}{
\begin{tikzpicture}
	\begin{pgfonlayer}{nodelayer}
		\node [style=none] (0) at (1, 2) {};
		\node [style=none] (1) at (1, -2) {};
		\node [style=none] (4) at (1, -1.75) {};
		\node [style=none] (5) at (1, -1.5) {};
		\node [style=none] (6) at (1, -1.25) {};
		\node [style=none] (7) at (1, -1) {};
		\node [style=none] (8) at (1, -0.75) {};
		\node [style=none] (9) at (1, -0.5) {};
		\node [style=none] (10) at (1, -0.25) {};
		\node [style=none] (11) at (1, 0) {};
		\node [style=none] (12) at (1, 0.25) {};
		\node [style=none] (13) at (1, 0.5) {};
		\node [style=none] (14) at (1, 0.75) {};
		\node [style=none] (15) at (1, 1) {};
		\node [style=none] (16) at (1, 1.25) {};
		\node [style=none] (17) at (1, 1.5) {};
		\node [style=none] (18) at (1, 1.75) {};
		\node [style=none] (19) at (1.5, -1.5) {};
		\node [style=none] (20) at (1.5, -1.25) {};
		\node [style=none] (21) at (1.5, -1) {};
		\node [style=none] (22) at (1.5, -0.75) {};
		\node [style=none] (23) at (1.5, -0.5) {};
		\node [style=none] (24) at (1.5, -0.25) {};
		\node [style=none] (25) at (1.5, 0) {};
		\node [style=none] (26) at (1.5, 0.25) {};
		\node [style=none] (27) at (1.5, 0.5) {};
		\node [style=none] (28) at (1.5, 0.75) {};
		\node [style=none] (29) at (1.5, 1) {};
		\node [style=none] (30) at (1.5, 1.25) {};
		\node [style=none] (31) at (1.5, 1.5) {};
		\node [style=none] (32) at (1.5, 1.75) {};
		\node [style=none] (33) at (1.5, 2) {};
		\node [style=none] (34) at (1.5, -1.75) {};
		\node [style=none] (35) at (1.5, -3) {$\rho=\infty$};
		\node [style=none] (36) at (2, 0) {\rotatebox{-90}{Boundary}};
		\node [style=none] (37) at (-0.5, 0) {Bulk};
		\node [style=SmallCircleRed] (38) at (1, 0.75) {};
		\node [style=SmallCircleRed] (39) at (1, -0.75) {};
		\node [style=none] (40) at (-1, 2) {};
		\node [style=none] (41) at (-1, 2.25) {};
		\node [style=none] (42) at (-1, -2) {};
		\node [style=none] (43) at (-1, -2.25) {};
		\node [style=none] (44) at (1, 2.25) {};
		\node [style=none] (45) at (1, -2.25) {};
		\node [style=none] (46) at (-1, -3) {$\rho<\infty$};
		\node [style=none] (47) at (-0.25, -1) {};
		\node [style=none] (48) at (-0.25, 1) {};
		\node [style=none] (49) at (-2, 2) {$\mathcal{D}_{\beta}$};
		\node [style=none] (50) at (-1.95, -2) {$\mathcal{D}^{\dagger}_{-\beta}$};
		\node [style=none] (51) at (8, 2) {};
		\node [style=none] (52) at (8, -2) {};
		\node [style=none] (53) at (8, -1.75) {};
		\node [style=none] (54) at (8, -1.5) {};
		\node [style=none] (55) at (8, -1.25) {};
		\node [style=none] (56) at (8, -1) {};
		\node [style=none] (57) at (8, -0.75) {};
		\node [style=none] (58) at (8, -0.5) {};
		\node [style=none] (59) at (8, -0.25) {};
		\node [style=none] (60) at (8, 0) {};
		\node [style=none] (61) at (8, 0.25) {};
		\node [style=none] (62) at (8, 0.5) {};
		\node [style=none] (63) at (8, 0.75) {};
		\node [style=none] (64) at (8, 1) {};
		\node [style=none] (65) at (8, 1.25) {};
		\node [style=none] (66) at (8, 1.5) {};
		\node [style=none] (67) at (8, 1.75) {};
		\node [style=none] (68) at (8.5, -1.5) {};
		\node [style=none] (69) at (8.5, -1.25) {};
		\node [style=none] (70) at (8.5, -1) {};
		\node [style=none] (71) at (8.5, -0.75) {};
		\node [style=none] (72) at (8.5, -0.5) {};
		\node [style=none] (73) at (8.5, -0.25) {};
		\node [style=none] (74) at (8.5, 0) {};
		\node [style=none] (75) at (8.5, 0.25) {};
		\node [style=none] (76) at (8.5, 0.5) {};
		\node [style=none] (77) at (8.5, 0.75) {};
		\node [style=none] (78) at (8.5, 1) {};
		\node [style=none] (79) at (8.5, 1.25) {};
		\node [style=none] (80) at (8.5, 1.5) {};
		\node [style=none] (81) at (8.5, 1.75) {};
		\node [style=none] (82) at (8.5, 2) {};
		\node [style=none] (83) at (8.5, -1.75) {};
		\node [style=SmallCircleRed] (87) at (8, 0.1) {};
		\node [style=SmallCircleRed] (88) at (8, -0.1) {};
		\node [style=none] (93) at (8, 2.25) {};
		\node [style=none] (94) at (8, -2.25) {};
		\node [style=none] (95) at (6, -0.1) {};
		\node [style=none] (96) at (6, 0.1) {};
		\node [style=none] (97) at (6, 2) {};
		\node [style=none] (98) at (6, 2.25) {};
		\node [style=none] (99) at (6, -2) {};
		\node [style=none] (100) at (6, -2.25) {};
		\node [style=none] (101) at (7, -0.1) {};
		\node [style=none] (102) at (7, 0.1) {};
		\node [style=none] (103) at (14, 2) {};
		\node [style=none] (104) at (14, -2) {};
		\node [style=none] (105) at (14, -1.75) {};
		\node [style=none] (106) at (14, -1.5) {};
		\node [style=none] (107) at (14, -1.25) {};
		\node [style=none] (108) at (14, -1) {};
		\node [style=none] (109) at (14, -0.75) {};
		\node [style=none] (110) at (14, -0.5) {};
		\node [style=none] (111) at (14, -0.25) {};
		\node [style=none] (112) at (14, 0) {};
		\node [style=none] (113) at (14, 0.25) {};
		\node [style=none] (114) at (14, 0.5) {};
		\node [style=none] (115) at (14, 0.75) {};
		\node [style=none] (116) at (14, 1) {};
		\node [style=none] (117) at (14, 1.25) {};
		\node [style=none] (118) at (14, 1.5) {};
		\node [style=none] (119) at (14, 1.75) {};
		\node [style=none] (120) at (14.5, -1.5) {};
		\node [style=none] (121) at (14.5, -1.25) {};
		\node [style=none] (122) at (14.5, -1) {};
		\node [style=none] (123) at (14.5, -0.75) {};
		\node [style=none] (124) at (14.5, -0.5) {};
		\node [style=none] (125) at (14.5, -0.25) {};
		\node [style=none] (126) at (14.5, 0) {};
		\node [style=none] (127) at (14.5, 0.25) {};
		\node [style=none] (128) at (14.5, 0.5) {};
		\node [style=none] (129) at (14.5, 0.75) {};
		\node [style=none] (130) at (14.5, 1) {};
		\node [style=none] (131) at (14.5, 1.25) {};
		\node [style=none] (132) at (14.5, 1.5) {};
		\node [style=none] (133) at (14.5, 1.75) {};
		\node [style=none] (134) at (14.5, 2) {};
		\node [style=none] (135) at (14.5, -1.75) {};
		\node [style=none] (138) at (14, 2.25) {};
		\node [style=none] (139) at (14, -2.25) {};
		\node [style=none] (142) at (12, 2) {};
		\node [style=none] (143) at (12, 2.25) {};
		\node [style=none] (144) at (12, -2) {};
		\node [style=none] (145) at (12, -2.25) {};
		\node [style=SmallCirclePurple] (146) at (14, 0) {};
		\node [style=SmallCircleGreen] (147) at (12, 0) {};
		\node [style=none] (148) at (12, 1) {};
		\node [style=none] (149) at (12, -1) {};
		\node [style=none] (150) at (0, -4.25) {(i)};
		\node [style=none] (151) at (7, -4.25) {(ii)};
		\node [style=none] (152) at (13, -4.25) {(iii)};
		\node [style=none] (153) at (11.5, 1) {$\mathcal{D}$};
		\node [style=none] (154) at (13, 0.5) {$\mathcal{C}$};
		\node [style=none] (155) at (7, -4.5) {};
	\end{pgfonlayer}
	\begin{pgfonlayer}{edgelayer}
		\draw [style=ThickLine] (0.center) to (1.center);
		\draw [style=ThickLine] (18.center) to (33.center);
		\draw [style=ThickLine] (17.center) to (32.center);
		\draw [style=ThickLine] (16.center) to (31.center);
		\draw [style=ThickLine] (15.center) to (30.center);
		\draw [style=ThickLine] (14.center) to (29.center);
		\draw [style=ThickLine] (13.center) to (28.center);
		\draw [style=ThickLine] (12.center) to (27.center);
		\draw [style=ThickLine] (11.center) to (26.center);
		\draw [style=ThickLine] (10.center) to (25.center);
		\draw [style=ThickLine] (9.center) to (24.center);
		\draw [style=ThickLine] (8.center) to (23.center);
		\draw [style=ThickLine] (7.center) to (22.center);
		\draw [style=ThickLine] (6.center) to (21.center);
		\draw [style=ThickLine] (5.center) to (20.center);
		\draw [style=ThickLine] (4.center) to (19.center);
		\draw [style=ThickLine] (1.center) to (34.center);
		\draw [style=DottedLine] (44.center) to (0.center);
		\draw [style=DottedLine] (1.center) to (45.center);
		\draw [style=DottedRed] (41.center) to (40.center);
		\draw [style=DottedRed] (43.center) to (42.center);
		\draw [style=RedLine, in=-180, out=-30, looseness=0.75] (48.center) to (38);
		\draw [style=RedLine, in=30, out=180, looseness=0.75] (39) to (47.center);
		\draw [style=ThickLine] (51.center) to (52.center);
		\draw [style=ThickLine] (67.center) to (82.center);
		\draw [style=ThickLine] (66.center) to (81.center);
		\draw [style=ThickLine] (65.center) to (80.center);
		\draw [style=ThickLine] (64.center) to (79.center);
		\draw [style=ThickLine] (63.center) to (78.center);
		\draw [style=ThickLine] (62.center) to (77.center);
		\draw [style=ThickLine] (61.center) to (76.center);
		\draw [style=ThickLine] (60.center) to (75.center);
		\draw [style=ThickLine] (59.center) to (74.center);
		\draw [style=ThickLine] (58.center) to (73.center);
		\draw [style=ThickLine] (57.center) to (72.center);
		\draw [style=ThickLine] (56.center) to (71.center);
		\draw [style=ThickLine] (55.center) to (70.center);
		\draw [style=ThickLine] (54.center) to (69.center);
		\draw [style=ThickLine] (53.center) to (68.center);
		\draw [style=ThickLine] (52.center) to (83.center);
		\draw [style=DottedLine] (93.center) to (51.center);
		\draw [style=DottedLine] (52.center) to (94.center);
		\draw [style=RedLine] (95.center) to (99.center);
		\draw [style=RedLine] (97.center) to (96.center);
		\draw [style=RedLine] (96.center) to (87);
		\draw [style=RedLine] (88) to (95.center);
		\draw [style=DottedRed] (100.center) to (99.center);
		\draw [style=DottedRed] (98.center) to (97.center);
		\draw [style=ArrowLineRed] (96.center) to (102.center);
		\draw [style=ArrowLineRed] (88) to (101.center);
		\draw [style=ThickLine] (103.center) to (104.center);
		\draw [style=ThickLine] (119.center) to (134.center);
		\draw [style=ThickLine] (118.center) to (133.center);
		\draw [style=ThickLine] (117.center) to (132.center);
		\draw [style=ThickLine] (116.center) to (131.center);
		\draw [style=ThickLine] (115.center) to (130.center);
		\draw [style=ThickLine] (114.center) to (129.center);
		\draw [style=ThickLine] (113.center) to (128.center);
		\draw [style=ThickLine] (112.center) to (127.center);
		\draw [style=ThickLine] (111.center) to (126.center);
		\draw [style=ThickLine] (110.center) to (125.center);
		\draw [style=ThickLine] (109.center) to (124.center);
		\draw [style=ThickLine] (108.center) to (123.center);
		\draw [style=ThickLine] (107.center) to (122.center);
		\draw [style=ThickLine] (106.center) to (121.center);
		\draw [style=ThickLine] (105.center) to (120.center);
		\draw [style=ThickLine] (104.center) to (135.center);
		\draw [style=DottedLine] (138.center) to (103.center);
		\draw [style=DottedLine] (104.center) to (139.center);
		\draw [style=DottedRed] (145.center) to (144.center);
		\draw [style=DottedRed] (143.center) to (142.center);
		\draw [style=BrownLine] (147) to (146);
		\draw [style=ArrowLineRed] (142.center) to (148.center);
		\draw [style=ArrowLineRed] (147) to (149.center);
		\draw [style=RedLine] (142.center) to (144.center);
		\draw [style=RedLine, in=150, out=-90] (40.center) to (48.center);
		\draw [style=RedLine, in=90, out=-150, looseness=0.75] (47.center) to (42.center);
		\draw [style=ArrowLineRed, in=150, out=-90] (40.center) to (48.center);
		\draw [style=ArrowLineRed, in=25, out=180, looseness=0.75] (39) to (47.center);
	\end{pgfonlayer}
\end{tikzpicture}
}
\caption{(i): Initial configuration of two defects $\mathcal{D}_{\beta}$ and $\mathcal{D}_{- \beta}^{\dagger}$ attaching to the boundary obtained by boosting with $\pm \beta$. (ii): Homotopically equivalent configuration to (i). (iii): We fuse two oppositely oriented segments of $\mathcal{D}_{\beta}$ and $\mathcal{D}_{-\beta}^{\dagger}$ resulting in the fusion product $\mathcal{C}$ and horizontally oriented defect $\mathcal{D}$ of the same dimension. The fusion product $\mathcal{C}$ can end on $\mathcal{D}$, which is the result of connecting $\mathcal{D}_{\beta}$  and $\mathcal{D}_{-\beta}^{\dagger}$ to a single object and will be associated with a symmetry operator.}
\label{fig:Fusion}
\end{figure}

As constructed, this defect still attaches to the
conformal boundary at $\rho\rightarrow\infty$. To detach it, we will consider a pair of such heavy defects. We boost one with boost parameter
+$\beta$ and one with boost parameter $-\beta$. In this case, since the pair
fuse at $\rho\rightarrow\infty$ we can now detach this defect at the expense of introducing a {fusion product $\mathcal{C}$ from oppositely oriented defects}
which stretches from the
bulk defect to the boundary CFT$_D$. The boundary of {$\mathcal{C}$}
fills the $X_{\text{new}}^{p}$ direction, and as $\beta\rightarrow\infty$ we
can push it arbitrarily close to the SymTFT sliver. In the
dual CFT, this is the combination of a dilatation and a rotation such that the
defect now fills an additional spatial direction. See figure \ref{fig:Fusion} for a depiction of this procedure.
A further comment is that in the full gravity dual we can expect this condensation defect $\mathcal{C}$ to dynamically
spread out as a flux tube which attaches the defect $\mathcal{D}$ back to the boundary CFT$_D$.

\subsection{Lorentzian AdS$_{D+1}$ Boosting}

We now consider a radially extended defect of Lorentzian AdS$_{D+1}$ for $D > 2$ and show how to push it fully into the SymTFT sliver. This can be used to build a class of symmetry operators for the boundary CFT$_D$.

To this end, consider the embedding space for Lorentzian AdS$_{D+1}$ in $\mathbb{R}^{2,D}$, as specified by the hypersurface:
\begin{equation}
    - L^{2}
    = -\left(  T^{1} \right)^{2} - \left(  T^{2} \right)^{2} + \left( X^{1} \right)^{2} + \dots + \left( X^{D} \right)^{2}.
\end{equation}
Global coordinates that satisfy this constraint equation are:
\begin{equation}
\begin{split}
    T^{1} &= L \cosh\rho \cos\tau \, ,
    \quad
    T^{2} = L \cosh\rho \sin\tau \, ,
    \\
    X^{j} &= L \Omega^{j} \sinh\rho \, ,
    \qquad
    \text{with }
    \,
    \underset{j=1}{\overset{D}{\sum}} \, \Omega^{j}\Omega^{j} = 1 \, ,
\end{split}
\end{equation}
in the usual notation.
Here, the constrained $\Omega^{j}$ parameterize an
$S^{D-1}$, and the radial coordinate $\rho \in [0, \infty)$ ranges from the deep interior ($\rho \to 0$) to the conformal boundary ($\rho \to \infty$).
We can adapt the present argument to more general boundary topologies so we leave this implicit in what follows.

Consider a heavy defect that sweeps out a codimension-$p$ subspace AdS$_{d+1}$, which without loss of generality we identify with setting $X^{1}=X^{2}=...=X^{p}=0$, where now we assume $p\geq2$.
We push this defect away from the deep interior by performing a pair of boosts in the $(T^{1},X^{p})$ and $(T^{2},X^{p-1})$ planes.
In terms of the boosted coordinates we have:\footnote{One can of course consider boosts by different amounts in the two planes, but the main idea is already established using the special choice considered here.}
\begin{align}
    \begin{bmatrix}
    T_{\text{new}}^{1}\\
    X_{\text{new}}^{p}
    \end{bmatrix}
    & =
    \begin{bmatrix}
    \cosh\beta & \sinh\beta\\
    \sinh\beta & \cosh\beta
    \end{bmatrix}
    \begin{bmatrix}
    T^{1}\\
    X^{p}
    \end{bmatrix} \, ,\\
    \begin{bmatrix}
    T_{\text{new}}^{2}\\
    X_{\text{new}}^{p-1}%
    \end{bmatrix}
    &=
    \begin{bmatrix}
    \cosh\beta & \sinh\beta\\
    \sinh\beta & \cosh\beta
    \end{bmatrix}
    \begin{bmatrix}
    T^{2}\\
    X^{p-1}
    \end{bmatrix}
\, .
\end{align}
In particular, since $X^{p-1}=X^{p}=0$ along the defect,  the defect is now (after boosting) localized along:
\begin{equation}
    X_{\text{new}}^{p} = T_{\text{new}}^{1} \tanh\beta
    \qquad
    \text{and}
    \qquad
    X_{\text{new}}^{p-1} = T_{\text{new}}^{2} \tanh\beta \, ,
\end{equation}
which in terms of the new global coordinates for Lorentzian AdS$_{D+1}$ is:
\begin{equation}
    \Omega^{p} \tanh\rho = \cos\tau \tanh\beta
    \qquad
    \text{and}
    \qquad
    \Omega^{p-1} \tanh\rho = \sin\tau \tanh\beta.
\end{equation}
Now, since $\left\vert \Omega^{p-1} \right\vert ^{2} + \left\vert \Omega^{p} \right\vert ^{2} \leq 1$, it follows that $\left\vert \tanh\rho \right\vert
\geq\tanh\beta$.
This inequality is saturated when $\left\vert \Omega^{p-1} \right\vert^{2} + \left\vert \Omega^{p}\right\vert^{2} = 1$.
In particular, this means that the defect has now been pushed to $\rho \geq \left\vert \beta \right\vert $.

Much as in the Euclidean signature case, to fully detach the defect from the boundary, we must also include another defect which is also boosted/rotated so that it smoothly matches onto the boundary profile.
In terms of global AdS$_{D+1}$ we have simply given our defect a large angular momentum and this angular momentum barrier prevents the object from falling too deep into the interior.

\subsection{The Special Case of AdS$_{3}$ and AdS$_{2}$}

In the previous subsections we gave a general procedure for pushing defects into the SymTFT sliver.
Some aspects of this discussion are different in the special case of AdS${_3}$ backgrounds so we now treat this case separately.
After this we briefly comment on the case of AdS${_2}$ backgrounds.

First, consider Euclidean AdS$_{3}$ with global coordinates
\be
\begin{aligned}
    T &= L \cosh\rho
     \\
    X^{1} &= L \sinh\rho \sin\theta \cos\phi
    \\
    X^{2} &= L \sinh\rho \sin\theta \sin\phi
    \\
    X^{3} &= L \sinh\rho \cos\theta \, .
\end{aligned}
\ee
Here, $\theta$ and $\phi$ are the usual polar and azimuthal angles respectively.
As mentioned earlier, the radial coordinate $\rho \in [0,\infty)$ ranges from the deep interior ($\rho \to 0$) to the conformal boundary ($\rho \to \infty$).

Consider a codimension-2 heavy defect that we identify by setting $X^{1} = X^{2} = 0$, or equivalently, $\theta = 0, \pi$.
Thus, the defect is given by
\begin{equation}
    T = L \cosh\rho
    \, ,
    \quad
    X^{3} = \pm L \sinh\rho \, .
\end{equation}
This defect stretches all the way from the deep interior to the conformal boundary.
As earlier, we can push this defect away from the deep interior by performing a boost in the $(T,X^{2})$ plane, and detach it from the conformal boundary by using a second defect with the opposite boost.

Next, consider Lorentzian AdS$_{3}$ with global coordinates
\begin{equation}
\begin{aligned}
    T^{1} &= L \cosh\rho \cos\tau \\
    T^{2} &= L \cosh\rho \sin\tau \\
    X^{1} &= L \sinh\rho \cos\phi \\
    X^{2} &= L \sinh\rho \sin\phi \, .
    \end{aligned}
\end{equation}
Here, $\phi$ is the usual azimuthal angle.
The codimension-2 heavy defect is identified by setting $X^{1} = X^{2} = 0$, or equivalently, $\rho = 0$.
Thus, the defect is given by
\begin{equation}
    T^{1} = L \cos\tau \, ,
    \quad
    T^{2} = L \sin\tau \, .
\end{equation}
Note that unlike the previous cases, this defect is localized at $\rho = 0$ in the deep interior.
Moreover, the defect is fully detached from the conformal boundary and we no longer need to boost it to localize it to a particular radial position. This defect can be sent closer to the conformal boundary by simply providing it with angular momentum.

Finally, let us briefly comment on the case of AdS$_2$ backgrounds. In this case the ``CFT$_{1}$'' is a 1D quantum mechanical system.\footnote{There is a subtlety in referring to this as a CFT since one now has a purely vanishing stress tensor. That being said, there is clearly a boundary system (suitably regulated) which captures many features of the bulk.} In this case, one can again proceed much as in the AdS$_3$ example for radially extended objects. Boosting such a defect into the boundary system will now result in a dimension-one object in the boundary system, i.e., it corresponds to the symmetry operator for a $(-1)$-form symmetry, which in turn amounts to varying a parameter of the boundary theory. This is especially interesting in the context of various ensemble averaged systems (see e.g., \cite{Maldacena:2016hyu, Heckman:2021vzx}).

\subsection{Worldvolume Theories and Boosting}

One can of course also apply the procedure in reverse, starting from a brane
at a fixed AdS radius close to the conformal boundary. Then, boosting it in the
same fashion will produce an object that extends radially in the AdS$_{D+1}$
directions.\footnote{Here we are discussing exclusively bulk configurations, e.g., the resulting heavy CFT defect can, but need not be, genuine. In particular, if the topological boundary conditions in the associated SymTFT\,/\,SymTh is such that the putative defect cannot end on the boundary we are necessarily constructing a relative defect. In this case, after boosting, we find a radially running bulk object which, upon reaching the topological boundary of the SymTFT\,/\,SymTh, where it can not end, continues to extend on within this boundary, however, now with the boundary conditions imposed along this segment of its worldvolume.} Now, these branes will come equipped with both a dynamical
fluctuating sector that depends on local perturbations of the metric and also topological contributions that are independent of such local metric perturbations.

In the limit where the brane
is pushed to the boundary, these topological components are the only surviving contributions to the topological symmetry operators.
For heavy defects, however, there is a priori no reason that such contributions have to decouple at all, and generically they do not.
Even so, we can still use our boosting formalism to determine the relation between the tension of the original heavy defect and its boosted counterpart.

In general terms, we can start with the stress-energy $T_{\mu\nu}^{\text{HD}}$
sourced by the heavy defect. Since we have an explicit coordinate transformation available, we can just boost this to the configuration
for the stress-energy $T_{\mu\nu}^{\text{B(HD)}}$. For this reason, it suffices to determine the tension of the heavy defect.

In the case of interest, we have a codimension-2 defect in the bulk. Strictly speaking, what we mean here is
that the semi-classical background contains a defect which asymptotes to a codimension-2 object in the boundary. In the deep bulk this defect attains finite thickness. Away from the AdS core, when this object
has uniform tension, we have:
\begin{equation}
\label{eq:Tension}
T_{\ast}^{\text{AdS-Defect}}=\frac{1}{\ell_{\ast}^{D-1}}=\frac{\chi}{8\pi
G_{D+1}},
\end{equation}
where $G_{D+1}$ is the Newton's constant in $D+1$ spacetime dimensions and $\ell_*$
is the characteristic length scale of the defect.
Here, $\chi$ is the conical deficit angle generated in the plane transverse to
the defect.

In subsequent sections, we will consider backgrounds of the form AdS$_{D+1} \times X$ as generated by the near horizon limit of branes probing Ricci-flat cones of the form $Y = \text{Cone} (X)$. In  AdS$_{D+1} \times X$, we then consider defects that deform the direct product structure. These will be of codimension-2 in the AdS$_{D+1}$ sliver, extend radially, and further exhibit deficit angles.
Some explicit truncated solutions with such features were recently presented in \cite{Arav:2024exg, Bomans:2024vii}.

Reduction of the extra-dimensional geometry results in a source of stress-energy which is concentrated in codimension-2. We comment, the defects we consider in AdS$_{D+1} \times X$ will be wrapped on internal loci $B$ of non-vanishing volume and it is theorefore appropriate to introduce an intrinsic tension $T_{\ast}^{\text{KK-Mag}}$ as set by the Kaluza-Klein reduction of the corresponding magnetic object (see \cite{Gross:1983hb}). After this reduction we are left with the AdS-Defect with tension \eqref{eq:Tension}. The two tensions are related by the volume of the wrapping locus
\begin{equation}
    T_{\ast}^{\text{AdS-Defect}} = T_{\ast}^{\text{KK-Mag}} \, \text{Vol}\,B  \, .
\end{equation}
Since there is typically no scale separation between the two factors of AdS$_{D+1} \times X_{m+1}$ there is an overall curvature lengthscale $L$ common to both factors.
Letting $\ell_{\text{PL}}$ denote the Planck length in $(D+m+2)$ dimensions, we then have $G_{D+1} \sim \ell_{\text{PL}}^{D+m}/L^{m+1}$
and the overall tension therefore goes as:
\begin{equation}
    T_{\ast}^{\text{AdS-Defect}} = \frac{1}{\ell_{\text{PL}}^{D-1}} \left(  \frac{L}{\ell_{\text{PL}}} \right)  ^{\gamma},
\end{equation}
for some $\gamma > 0$.\footnote{ Comparing this result with \eqref{eq:Tension}, we note that the additional dependence on $\frac{L}{\ell_{\text{PL}}}$ comes from the conical deficit $\chi$  not necessarily being $O(1)$.}
This is in accord with the general considerations presented in \cite{Heckman:2024oot}.

\section{Defects, Symmetry Operators, and Isometries} \label{sec:DEFECTS}

In this section, we turn to the gravity duals of CFT$_D$ continuous symmetry operators which descend from internal isometries of AdS$_{D+1}\times X$.
From the perspective of gravity on AdS$_{D+1}$, the isometries of $X$ correspond to a gauge theory with standard vector potentials.
We label the continuous symmetry group as $G$.
Proposals for the SymTFT\,/\,SymTh of a continuous symmetry have been put forward in \cite{Heckman:2024oot, Brennan:2024fgj, Antinucci:2024zjp, Bonetti:2024cjk, Apruzzi:2024htg, Argurio:2024oym, Antinucci:2024bcm, Cvetic:2024dzu}.
In principle, we can consider either an abelian or a non-abelian gauge group, but for ease of exposition, we focus on the special case $G = U(1)$ since the other cases can be handled similarly.

The consistent truncation procedure (see, e.g., \cite{Cvetic:2000dm, Cvetic:2000nc}) tells us that the gravitational theory will include a gauge
theory sector
\begin{equation}
    \mathcal{L}_{\text{AdS}} \supset - \frac{1}{2g^{2}} F \wedge \ast F,
\end{equation}
in the usual notation. We focus on CFT$_D$'s with $D\geq 3$; when $D+1\leq3$ the kinetic term will be dominated at
long distances by Chern-Simons-like interaction terms. See, e.g., \cite{Montero:2016tif} for a helpful discussion and review on this point.
Further, for ease of exposition, we suppress bulk Chern-Simons terms and other interaction terms as they can easily be reincorporated and do not affect the main elements of our proposal.\footnote{Of course, these Chern-Simons terms are quite important in determining the worldvolume theory of the extended objects we construct which interact with these via inflow. See \cite{Bah:2023ymy, Waddleton:2024iiv} for related analyses involving Page charges. We will return to these issues when concerned with the defect worldvolume theory.}

In this section considerations will be restricted to the asymptotic AdS sliver hosting the SymTFT\,/\,SymTh. There, $F$ is flat due to the formally infinite volume, and as a device to restrict to flat field configurations, one can introduce a Lagrange multiplier / BF term\footnote{We note that this extends to the case where $G$ is non-abelian by writing $\Tr (C_{D-1} \wedge F)$.
Here, the $(D-1)$-form potential $C_{D-1}$ takes values in the non-abelian Lie algebra.
One might worry here about constructing a suitable path
ordered exponentials for this non-abelian potential.
In the present setting where one restricts to flat gauge field configurations, one can only label these defects by conjugacy classes of $G$ rather than actual group elements.
See, e.g., \cite{Bonetti:2024cjk} for further discussion on this point.}
\begin{equation}\label{eq:FlatnessForce}
    \mathcal{L}_{ \text{SymTFT}} \supset \frac{1}{2\pi} C_{D-1} \wedge F \, ,
\end{equation}
Observe that the equation of motion for the gauge field now yields:
\begin{equation}
\label{eq:dFdual}
    \dd F_{D-1}^{\text{dual}} + \dots = \frac{g^2}{2\pi} \, \dd C_{D-1} \, ,
\end{equation}
where $F_{D-1}^{\text{dual}} = \ast F_{2}$ is the magnetic dual field strength.
As a general comment, the ``...'' refers to
additional terms in the equation of motion (such as those coming from possible Chern-Simons-like terms).

Concentrating on the SymTFT\,/\,SymTh sliver, we need to find candidate gravity duals for the pair of operators:
\begin{equation}
    W_{n}=\exp \left( i n\int A_{1} \right)
    \qquad
    \text{and}
    \qquad
    T_{\alpha} = \exp \left(  i \alpha \int C_{D-1} \right) \,.
\end{equation}
Here the gauge field $A_{1}$ is $U(1)$-valued, the Lagrange multiplier $C_{D-1}$ is $\mathbb{R}$-valued,
and we have parameters $n \in \mathbb{Z}$ and $\alpha \in [0,1)$. Using the equation of motion in \eqref{eq:dFdual}, observe that $C_{D-1}$ is simply $\frac{2\pi}{g^2} F_{D-1}^{\text{dual}}$. As such, we can build these operators not only in the SymTFT but also in the
gravitational system by using $F_{D-1}^{\text{dual}}$ instead of $C_{D-1}$.

At this point, we encounter an important subtlety in comparing the gravity approach with that of the SymTFT.
It is widely expected that in any consistent theory of quantum gravity, the only available continuous gauge groups are compact.
In particular, we expect that $F_{D-1}^{\text{dual}}$ is
valued in $U(1)$ rather than in $\mathbb{R}$, which is where the SymTFT field $C_{D-1}$ takes values.
It is here that treating the SymTFT sector as a small sliver of the gravity dual becomes quite helpful.
Observe that we can detect quantization of $F_{D-1}^{\text{dual}}$ by integrating over compact subspaces in the bulk.
However, if we restrict to a small patch (the SymTFT sliver) then this quantization condition is destroyed.
The parameter $\alpha$ can be viewed as a delta function supported improperly quantized background flux, so in the SymTFT sliver we have
\begin{equation}\label{eq:fluxalpha}
    T_{\alpha} = \exp\left(  \frac{i}{2 \pi} \underset{(D+1)\text{-Sliver}}{\int} F_{2}^{\text{bkgnd}} \wedge C_{D-1} \right)  = \exp \left( i \alpha \underset{Y_{D-1}}{\int} C_{D-1} \right)  ,
\end{equation}
in the usual notation.  Here $\alpha Y_{D-1}$ is the Poincar\'e dual to the background 2-cocycle $F_2^{\text{bkgnd}}$. Again, integrating $F_{2}^{\text{bkgnd}}$ over compact subspaces would have resulted in a quantized answer, but the restriction to non-compact subspaces in the SymTFT sliver allows us to have more general field valuations.

Let us now turn to possible boundary conditions for our SymTFT sliver.
There are two canonical choices for the boundary conditions one might entertain, see, e.g., \cite{Brennan:2024fgj}.
One choice is to allow the Wilson line $W_{n}$ to extend from $\left\vert \text{top} \right\rangle$ to $\left\vert \text{phys} \right\rangle$, i.e., it specifies a heavy defect.
In the CFT$_{D}$, the endpoint of this line specifies a pointlike object, i.e., a local operator.
In this case, the $T_{\alpha}$'s play the role of the symmetry operators, and indeed, they are codimension-1 operators inside of the CFT$_{D}$.

Another choice is to allow the $T_{\alpha}$ to extend from $\left\vert
\text{top}\right\rangle $ to $\left\vert \text{phys}\right\rangle $ as associated with a heavy defect.
In the boundary CFT$_{D}$ this is interpreted as a codimension-2 operator, and is labeled by the continuous parameter $\alpha$.\footnote{In the case of a non-abelian symmetry group $G$, we would label this operator by a choice of conjugacy class $[g]\in$ Conj $G$, if we aim to ultimately realize a genuine operator.}
For this choice of polarization, the $W_{n}$'s play the role of the symmetry operators that link the charged objects.

We remark that in what follows we shall mainly focus on the electric polarization, i.e., where the $T_{\alpha}$ specify genuine symmetry operators. Indeed, the CFT$_{D}$ generically has continuous symmetry anomalies that dictate, for instance, the R-symmetry (and other) anomalies of the system. Thus, gauging this symmetry cannot be done and we are stuck in the electric polarization.

The rest of this section is organized as follows. We begin by sketching how we will use the boosting procedure of section \ref{sec:BOOST} to
build a radially extended defect. In the CFT$_D$ as well as the dual AdS$_{D+1}$ this will be associated with 2-form flux which is interpreted as sourcing a codimension-2 defect. Our aim will be to lift this object to AdS$_{D+1} \times X$, where we shall interpret the specific flux background as a choice of higher-dimensional metric / field configuration. To work up to this we consider an intermediate reduction where we explicitly identify the KK gauge boson via various intermediate reductions and consistent truncations. The gauge theory data of the CFT$_D$ / flux data of the AdS$_{D+1}$ can then be lifted to a topological fibration of $X$ over the AdS$_{D+1}$ base. Thankfully, even more is known about these fibrations in special cases, e.g., the prototypical case of AdS$_5 \times S^5$ \cite{Arav:2024exg, Bomans:2024vii} where the full metric and background field profiles is already known! In particular, we use this additional information to read off the tension for the bulk duals to our symmetry operators. We also discuss some generalizations to more general backgrounds of the form AdS$_{D+1} \times X$.

\subsection{Building Defects via Flux Data} \label{ssec:FLUXDATA}

In this section we sketch in broader terms our strategy for building the gravity duals to the desired symmetry operators $T_{\alpha}$.
Our starting point is equation (\ref{eq:fluxalpha}) which shows that in the SymTFT sliver we can build the desired operator provided we have switched on a suitable background flux. We shall be interested in a concentrated flux profile and we shall refer to this as a ``fluxbrane''.\footnote{Indeed, from the perspective of building the topological operators for a continuous symmetry we should expect fluxbranes to appear, as in reference \cite{Cvetic:2023plv}. A further comment is that many fluxbrane configurations can be explicitly realized via a brane / anti-brane annihilation process (see the Appendix of \cite{Cvetic:2023plv}). See also the later discussion given in \cite{Bergman:2024aly} for a related treatment in terms of non-BPS branes as well as \cite{Dierigl:2023jdp} for an earlier discussion indicating the use of other sorts of non-BPS branes as symmetry operators.}

Since we shall be using our boosting procedure to take radially extended objects and pushing them into the SymTFT sliver, it is natural to begin in the CFT$_D$ by building a codimension-2 flux defect. In the gravity dual this flux defect extends out radially. We shall then boost this into the boundary theory to build the desired codimension-1 symmetry operator. Along these lines, we shall be interested in a delta function supported flux associated with a choice of background field configuration for our global symmetry $G$. Denoting by $f_{2}$ the field strength and $a$ the local gauge connection, we can build the desired flux profile from a codimension-2 flux defect of the form:
\begin{equation}
a = \frac{1}{i} h^{-1} \dd h,
\end{equation}
where here, we have introduced a singular gauge transformation via $h = \exp(i \alpha \phi)$ with $\alpha \in \mathfrak{g}$ our fixed element
in the Lie algebra, and $\phi$ an angular coordinate which winds around the codimension-2 defect of the CFT$_D$. Observe that the resulting flux for this defect is then of the desired form:\footnote{As
explained in \cite{Gukov:2006jk}, this is a bit imprecise (but will suffice for our present considerations)
since shifts in $\alpha$ by elements of the root lattice for $G$ ought to produce the same field
strength, something which clearly will not happen in the present setting.
Reference \cite{Gukov:2006jk} give a more precise definition in terms of boundary values
specified by a corresponding conjugacy class for a group element.}
\begin{equation}
f_2 = 2 \pi \alpha \delta^{(2)}(\Sigma_{D-2}).
\end{equation}
More precisely, we can simply work in terms of a singular gauge field configuration as labelled by Gukov-Witten operator associated with the background global symmetry $G$.

We are especially interested in cases where the flux is \textit{not} properly quantized since we need to produce a general element of the global symmetry $G$. This is to be expected since radially extending this flux defect into the SymTFT sliver extends this to a codimension-2 flux defect $F_{2}^{\mathrm{bkgnd}}$ which is \textit{also} improperly quantized (see equation (\ref{eq:fluxalpha})). This is acceptable since we are localizing the flux on a non-compact subspace with boundary.

Having specified a radially dependent flux profile, our goal will be to geometrize this choice of background flux $F_{2}^{\mathrm{bkgnd}}$. On general grounds, these gauge fields / fluxes arise from the metric isometries of the higher-dimensional space $X$. As such, we expect the choice of gauge field to lift to a metric fibration of $X$ over AdS$_{D+1}$ and a non-trivial flux to be specified by a choice of curvature. We work up to this picture by considering various reductions / truncations of the higher-dimensional background geometry. With this in place, we can then apply our boosting procedure from section \ref{sec:BOOST} to build a defect localized in the SymTFT$_{D+1}$ sliver which in turn furnishes the dual of the continuous topological symmetry operator.

\subsection{Intermediate Reductions \label{ssec:REDUCTION}}

We now turn to discuss geometric properties of the AdS bulk dual $T_\alpha$ of the CFT symmetry operators. We begin by focusing on topological features, and initially consider the case related to a $U(1)$ isometry. For ease of exposition, we focus on the case where we have a 10D spacetime of the form AdS$_{D+1} \times X$, but clearly the considerations we present apply more broadly.

To begin, we consider defect insertions that deform the direct product structure of AdS$_{D+1}\times X$. Away from the defect locus, we will still have fibers $X$, but these now fiber non-trivially over spheres linking the defect and are permitted to degenerate along the defect locus. In the cases we consider $X$ is only twisted along some of its directions $F$, and whenever these can be isolated by a (possibly degenerate) fibration $F\hookrightarrow X\rightarrow B$ then we will analyze this twisting via an intermediate reduction to $B$.

Now, focus on a specific $U(1)$ isometry subgroup. When the isometry action is fixed point free we have $F=S^1$ and $B$ is closed. When the isometry acts with fixed points in codimension-2 then generically $F=S^1$ with exceptional fibers where the circle has pinched to points and $B$ is a manifold with boundary. Fixed points in higher codimension can result in more general, singular bases $B$. Further, discrete subgroups $\mathbb{Z}_M\subset U(1)$ can also have fixed points, in this case the base $B$ will contain quotient singularities.

Consider for example the fixed point free 5-dimensional case, to which we associate the non-degenerate $S^{1}$ fibration $S^{1}\hookrightarrow X_{5} \rightarrow B_{4}$ over some 4-dimensional compact base $B_{4}$. Then, we can consider the formal reduction along the $U(1)$ isometry to reach a 9D spacetime AdS$_{5}\times B_{4}$.
In this 9D spacetime, we have an electric vector potential $A_{1}$ with field strength $F_{2}$, and its magnetic dual $A_{6}$ with field strength $F_{7}$. On AdS$_5\times B_4$ we are now looking for backgrounds satisfying
\begin{equation}
\label{eq:alpha2}
    F_2 = 2 \pi \alpha \, \delta^{(2)}(\Sigma_7)\,,
\end{equation}
in the sliver (following the general discussion near line \eqref{eq:fluxalpha}), and which are (away from $\Sigma_7$) flat codimension-2 configurations with holonomy $\alpha$ along linking paths. Such objects are referred to as fluxbranes \cite{Gutperle:2001mb, Emparan:2001gm, Cvetic:2023plv} as they electrically couple to the field strength $F_7=\dd A_6$.

\subsection{Topology of the Lifted Fluxbrane}
\label{ssec:TopCon}

Having characterized the flux profile associated with a radially extended fluxbrane configuration, our aim will now be to lift this configuration back into pure geometry. To be concrete we illustrate these considerations by focusing on type IIB backgrounds of the form AdS$_{5} \times X_5$, but we again emphasize that these considerations apply more generally.

Due to the fact that we have a flux defect, we now ask how AdS$_5\times X_5$ can be degenerated and twisted to contain a codimension-2 object in AdS$_5$ which realizes a monodromy action on $X_5$ belonging to a $U(1)$ subgroup of $G=\text{Isom}(X_5)$. We restrict our considerations for now to the asymptotic SymTFT\,/\,SymTh sliver of the $\text{AdS}_5$ where the flux can be localized to a brane.

\begin{figure}
\centering
\scalebox{0.8}{
\begin{tikzpicture}
	\begin{pgfonlayer}{nodelayer}
		\node [style=none] (23) at (2.5, -2) {};
		\node [style=none] (24) at (5, 0.5) {};
		\node [style=none] (25) at (7, -2) {};
		\node [style=none] (26) at (9.5, 0.5) {};
		\node [style=none] (28) at (6.5, -0.25) {};
		\node [style=none] (29) at (5.5, -0.25) {};
		\node [style=none] (30) at (5.5, -1.25) {};
		\node [style=none] (31) at (6.5, -1.25) {};
		\node [style=NodeCross] (32) at (6.75, -0.75) {};
		\node [style=none] (33) at (6.75, -0.5) {};
		\node [style=none] (34) at (7.5, 1.25) {};
		\node [style=none] (35) at (7.75, 1.75) {$X_5\mapsto g\cdot X_5$};
		\node [style=Star] (37) at (6, -0.75) {};
		\node [style=none] (38) at (8.75, -1.25) {$V_5^g$};
		\node [style=none] (39) at (6, -0.5) {};
		\node [style=none] (40) at (5.25, 1.25) {};
		\node [style=none] (41) at (5, 1.75) {$X_5/U(1)_g$};
		%\node [style=none] (42) at (4.875, -0.75) {$k_*$};
		\node [style=none] (42) at (5, -2.75) {};
	\end{pgfonlayer}
	\begin{pgfonlayer}{edgelayer}
		\draw [style=ThickLine] (24.center) to (26.center);
		\draw [style=ThickLine] (26.center) to (25.center);
		\draw [style=ThickLine] (25.center) to (23.center);
		\draw [style=ThickLine] (23.center) to (24.center);
		\draw [style=ThickLine, bend right=45] (28.center) to (29.center);
		\draw [style=ThickLine, bend right=45] (29.center) to (30.center);
		\draw [style=ArrowLineRight, bend right=45] (30.center) to (31.center);
		\draw [style=DottedLine] (34.center) to (33.center);
		\draw [style=DottedLine] (40.center) to (39.center);
	\end{pgfonlayer}
\end{tikzpicture}
}
\caption{Sketch of the local model for the codimension-2 isometry defect. We show the internal space as a fibration over the two dimensions of $ V^g_5$ transverse to the defect. The fluxbrane wraps the 4D degenerated fiber collapsing the KK circle there and extends in three additional AdS dimensions. The internal space $X_5$ undergoes monodromy $X_5\rightarrow g\cdot X_5$ along loops linking the defect. The deficit angle of the cone $V^g_5$ is determined by the total flux, which, modulo some periodicity, also determines the monodromy rotation in $U(1)_g$. }
\label{Fig:localGeo}
\end{figure}
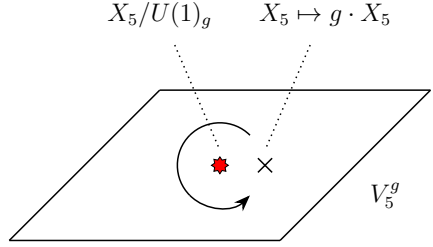

To begin, consider type IIB on AdS$_{5}\times X_5$ as obtained from the near horizon limit of $N$ D3-branes probing the tip of the Calabi-Yau cone $\text{Cone}(X_5) = Y$.
Let $G$ denote the isometry group of $X_5$, and consider a specific group element $g \in G$ which is associated with a Killing vector $\xi_{g}$ of $X_5$ and sweeps circles $S^{1}_g=U(1)_g$ in $X_5$.
Without loss of generality, we can then parameterize elements on $S^1_g$ following
\begin{equation}\label{eq:gt}
    g \left(  \alpha \right)  = \exp(2 \pi i \alpha t) \, ,
\end{equation}
where $t \in \mathfrak{g}$ is a generator of the Lie algebra, and $\alpha \in [0,1)$.

Denote by $U_5\setminus \Sigma_3$ a local patch in AdS$_5$ with some codimension-2 locus deleted. If $\Sigma_3$ is to support a localized defect associated with a isometry rotation by $g$ then we have
\be\label{eq:Mono}
\text{Monodromy}\,:\qquad X_5\rightarrow g\cdot X_5\,,
\ee
along any loop $\sigma_1$ linking $\Sigma_3$. The exceptional fiber topological consistent with this generic monodromy and projecting onto the defect locus is
\be\label{eq:Degen}
\text{Exceptional Fiber}\,:\qquad X_5/U(1)_g\,,
\ee
where $U(1)_g=\{\:\!g(\alpha)\,|\,\alpha \in [0,1)\}$. Observe that when $g$ is of finite order, a quotient by a finite group of that order, such that the exceptional fiber remains a five-dimensional space, is also consistent. In the generic situation, however, the quotient actually leads to a lower-dimensional space. Indeed, we have simply collapsed a circle, in similar fashion to the pinching of circles in the multi-centered Taub-NUT metric. But unlike that case (where $g$ is always of finite order), which describes a monopole configuration of codimension-3, we have pinched the circle $U(1)_g$ along a codimension-2 locus achieving a vortex-like configuration.

Overall the defect insertion therefore describes a deformation of the direct product $U_5\times X_5$ to a space projecting onto $V^g_5$ (which is $U_5$ with a deficit angle, but topologically unaltered otherwise) with generic fiber $X_5$, exceptional fiber $X_5/U(1)_g$ and the prescribed monodromy $g$. See figure \ref{Fig:localGeo}. Via \eqref{eq:gt} the initial group element $g$ can be thought of as the Lie algebra direction $t/|t|$ and its magnitude $|t|$. The space $X_5/U(1)_g$ is determined from the direction $t/|t|$ whereas the magnitude $|t|$ (modulo some period) maps onto the monodromy. Further, the magnitude $|t|$ (without any identifications) also specifies the deficit angle of $V_g$.

\begin{figure}
\centering
\scalebox{0.8}{
\begin{tikzpicture}
	\begin{pgfonlayer}{nodelayer}
		\node [style=none] (0) at (-2, 1.5) {};
		\node [style=none] (1) at (-2, -1.5) {};
		\node [style=none] (2) at (2, 1.5) {};
		\node [style=none] (3) at (2, -1.5) {};
		\node [style=Star] (4) at (0, 0) {};
		\node [style=none] (5) at (2, 0) {};
		\node [style=none] (6) at (-3, -1) {};
		\node [style=none] (7) at (-3, 1) {};
		\node [style=none] (8) at (-1, -2.25) {};
		\node [style=none] (9) at (1, -2.25) {};
		\node [style=none] (10) at (-3.5, 0) {$x_\perp$};
		\node [style=none] (11) at (0, -2.75) {$r$};
		\node [style=none] (12) at (-2, 2) {};
		\node [style=none] (13) at (2, 2) {};
		\node [style=none] (14) at (2, -2) {};
		\node [style=none] (15) at (-2, -2) {};
		\node [style=none] (16) at (0, -0.75) {$\Sigma_3$};
		\node [style=none] (17) at (4, 0) {};
		\node [style=none] (18) at (0, 2.625) {SymTFT / SymTh};
		\node [style=NodeCross] (19) at (-2, 0) {};
		\node [style=none] (20) at (3.5, 0.25) {Topological};
		\node [style=none] (21) at (3.5, -0.25) {Boundary};
	\end{pgfonlayer}
	\begin{pgfonlayer}{edgelayer}
		\draw [style=ThickLine] (0.center) to (1.center);
		\draw [style=ThickLine] (2.center) to (3.center);
		\draw [style=ArrowLineRight] (8.center) to (9.center);
		\draw [style=ArrowLineRight] (6.center) to (7.center);
		\draw [style=DashedLine] (4) to (19.center);
		\draw [style=DottedLine] (12.center) to (0.center);
		\draw [style=DottedLine] (13.center) to (2.center);
		\draw [style=DottedLine] (14.center) to (3.center);
		\draw [style=DottedLine] (15.center) to (1.center);
	\end{pgfonlayer}
\end{tikzpicture}}
\caption{Sketch of the codimension-2 R-symmetry defect in the SymTFT\,/\,SymTh sliver. The defect is supported on $\Sigma_3$ away from the physical boundary, however, it is connected thereto by a monodromy branch cut. In the dual CFT, the endpoint of this branch cut ($\times$) realizes the codimension-1 R-symmetry defect.}
\label{fig:DualityDefectLike}
\end{figure}
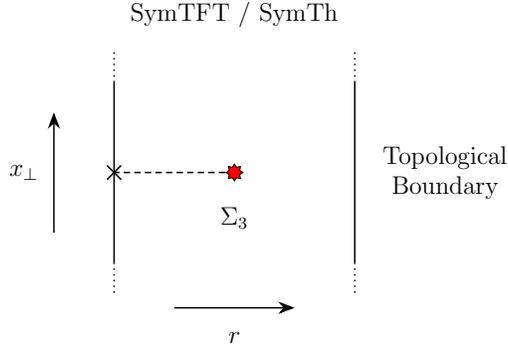

Notice that the constructed defects are monodromy defects in the sense of \cite{Heckman:2022xgu, Kaidi:2022cpf,Arav:2024exg}. Denoting the two coordinates of the sliver transverse to $\Sigma_3$ by $r,x_\perp$ we can, after projecting onto the topological structures, concentrate the monodromy to a branch cut running from the defect to the physical boundary condition. While such considerations are accurate in the SymTFT\,/\,SymTh sliver, they will need to be revisited away from this in the AdS bulk. Very similar to \cite{Heckman:2022xgu} the branch cut can in principle support terms of its own, however when oriented as displayed in figure \ref{fig:DualityDefectLike}, collapsing the SymTFT\,/\,SymTh slab in the horizontal direction renders them inconsequential in the CFT dual, and we therefore do not discuss these further.

So far we have focused on the topological profile of a single monodromy defect. Of course, it is also important to study the effects of bringing more than one defect together. Along these lines, we now consider two elements $g,g' \in G$ and their corresponding defects $D_{g}$ and $D_{g'}$. In the group, we know that the product returns a third element $g'' = g g' \in G$, so it is natural to expect the same to hold for the accompanying defects. In principle there can be a more intricate fusion rule due to the worldvolume dynamics of the brane, an issue we defer to section \ref{sec:Bulk}. With this in mind, we need to check that the product of our two defects has the general form:
\begin{equation}
D_g\otimes D_{g'}=D_{gg'}+ ...
\end{equation}
where the ``...'' indicate additional terms stemming from possible non-invertiblity. Notice that earlier we have simply specified a metric monodromy defect, as parameterized by $g\in G$, and have not discussed, for example, any other degrees of freedom of the associated background. In this sense our discussion was universal. Indeed, to determine additional degrees of freedom supported on defects we would have to delve into the supergravity theory we are considering on the above spaces, which we defer to section \ref{sec:PROPERTIES}. As such, here, we will only track the geometric ``top charge'' (i.e., the coupling to $F_7$) of the defect brane across fusion.

With this established, note that the desired fusion for the case $g'\in U(1)_g$ is immediate. Consider next the case where $U(1)_g,U(1)_{g'}$ are distinct and apply the intermediate reduction procedure of section \ref{ssec:REDUCTION} twice. We are therefore viewing $X_5\rightarrow \tilde{B}_3$ as a fibration over some 3-dimensional base $\tilde{B}_3$ with generic fiber
\be
T^2_{g,g'}=S^1_{g}\times S^1_{g'}\,.
\ee
When the defects are separated,  $S^1_{g}$ and $S^1_{g'}$ collapse separately at $D_{g}$ and $D_{g'}$, respectively. This specifies a distinguished $\mathfrak{u}(1)^2$ inside of $\mathfrak{g}$ and as such all charges are again elements in this distinguished direction. In particular, the ellipses do not contain any fusion products of the form $D_{k}$ for some $k \neq g g'$. This is because we have already determined the exceptional fiber projecting to the defect locus resulting from fusion to $X_5/U(1)_{gg'}$.

Finally, we note that the same considerations generalizes to other vacua of the form AdS$_{D+1} \times X$ for general $D > 2$. In the cases $D = 1,2 $ we comment that the only subtlety we need to contend with is the IR dynamics of the associated bulk gauge theory $G$, but at the level of classical background geometries everything goes through as above.

\subsection{Boundaries and Singularities of Fluxbranes}
\label{ssec:FluxBC}

We now discuss the boundaries of a fluxbrane in the SymTFT\,/\,SymTh sliver. The fluxbrane itself is topologically characterized by a collection of exceptional fibers $X_5/U(1)=B_4$. The fluxbrane wraps $\Sigma_3\times B_4$ and so whenever $B_4$ is not closed and smooth we need to specify additional data along the boundaries / irregular loci. The topology of $B_4$ is determined by the fixed point structure of the $U(1)$ action. For example, if fixed points occur in codimension-2 then $B_4$ has a boundary $B_3=\partial B_4$. More generally, we have singularities whenever fixed points occur in higher codimension or subgroups of $U(1)$ have fixed points.

Consider for example the case $X_5=S^5$ acted on by $U(1)\subset \mathrm{Isom}(S^5)$. Then fixed point loci have even codimension.\footnote{This follows straightforwardly by considering the action of $U(1)$ on the tangent bundle $T_pS^5$ at a fixed point $p$. This action diagonalizes, with complex unit norm eigenvalues. Non-real eigenvalues come in conjugate pairs and the eigenvalue 1 appears an odd number of times. The latter indicates tangent bundle directions from which we can generate a flow to sweep out the fixed point locus. } They are either empty or $S^1, S^3,S^5$ with the last case covering the trivial action. With this diagonal actions of type $(z_1,z_2,z_3)\mapsto (e^{im_1\theta} z_1,e^{im_2\theta} z_2, e^{im_3\theta}z_3)$ on $\mathbb{C}^3 = \mathrm{Cone}(S^5)$ already provide representative examples for respectively setting none, one, two, or three of the $m_i$ to zero.\footnote{Even when all $m_i$ are non-vanishing subgroups of $U(1)$ can have fixed points. Indeed, the quotient $S^5/U(1)=\mathbb{WCP}^2_{m_1,m_2,m_3}$ is a weighted projective space, and the three affine patches display quotient singularities modeled on $\mathbb{C}^2/\mathbb{Z}_{m_k}$ with weights $m_i,m_j$ where $\{i,j,k\}=\{1,2,3\}$.}

Consider now more closely the cases with fixed point locus either $S^1$ or $S^3$, respectively. Here, we can think of the $S^5$ as fibered over the ball $\mathbb{B}^4$ or $\mathbb{B}^2$, respectively. The fibers are $S^1$ and $S^3$ respectively, which are not acted on. We then see $S^5/U(1)$ is an $S^1$ fibration over $\mathbb{B}^3$ or an $S^3$ fibration over $[0,1]$. The second case, for example, is simply a collection of $S^3$'s with radius $r\in [0,1]$. In the former case, there is a singularity. In both cases, the geometries fiber non-trivially over the two AdS directions normal to the bulk defect and combine with the deficit angle into the total relevant singular geometry.

Let us focus now on cases with at most codimension-2 fixed point loci such that $B_4$ is a smooth manifold with or without boundary. The fluxbrane wraps $\Sigma_3\times B_4$ with boundary
\be
\partial (\Sigma_3\times B_4)=(\partial \Sigma_3)\times B_4 \,\sqcup\, \Sigma_3\times( \partial B_4)\,.
\ee
Whenever the associated $U(1)$ action is fixed point free we have $\partial B_4=0$. When fixed point loci occur in codimension-2 we will assume $\partial \Sigma_3=0$ (without this assumption $\Sigma_7=\Sigma_3\times B_4$ develops corners and we would need to additionally characterize how the boundary conditions along the two boundary components interact at the corner).

In both cases, our boundary conditions of choice at $\partial \Sigma_7$  are 10D KK 5-branes, which from the perspective of the 9D intermediate reduction spacetime $\text{AdS}_5\times B_4$ are monopole 5-branes of the circle reduction gauge field. In the sliver, like the fluxbranes, they uplift to singular metric profiles with localized topological worldvolume degrees of freedom constrained by anomaly inflow from the fluxbrane worldvolume. Note however, that whenever considering stacks of  $K\in \mathbb{Z}_{\geq 0}$ KK 5-branes the rotation angle $\alpha$ in \eqref{eq:alpha2} and \eqref{eq:gt} is quantized. Realizing more general values of $\alpha$ simply means we are dealing with unquantized fluxes, and much as in other contexts this will be referred to as ``fluxbranes''.

Codimension-4 fixed point loci with $\partial \Sigma =0$ are best understood via an intermediate reduction to 8D with two circle fibers, and two associated gauge fields $A_1^{(I)}$. Singularities of $S^5/U(1)$ then electrically couple to both the electromagnetically dual 5-form connections ${A}_5^{(I)}$. Singularities that occur due to subgroups of $U(1)$ having fixed points may be analyzed similarly, however, such singularities are already readily interpreted in 9D.\footnote{For example, consider $S^5/U(1)=\mathbb{WCP}^2_{m_1,m_2,m_3}$ with $\text{gcd}(m_{i},m_j)=m_{ij}$. Then, when $m_{ij}>1$, the quotient singularities can be interpreted as intersections of KK 5-branes.} In general terms, the overall geometry features a singularity which then dictates how the degrees of freedom localized to the monodromy defect interact with the CFT$_4$.\footnote{Compare this with the construction of duality defect in \cite{Heckman:2022xgu}, where the defect is constructed from a non-perturbative 7-brane. There, ultimately the interaction with the 4D QFT reduced to 4D backgrounds coupling to the minimal abelian TFT of reference \cite{Hsin:2018vcg}.}

We now discuss the topological features of these combined KK 5-brane and fluxbrane configurations when pushing these from the sliver into the physical AdS bulk at a finite AdS radius. After such a push the flux spreads out and can no longer be localized to $\Sigma_7$, as this flux is abelian. We can argue for this already in the 10D IIB supergravity approximation of the setup. In the sliver, the internal space $X_5$ is treated as having formally infinite volume. As a consequence, all supergravity fluctuations with positive definite kinetic terms are such that these exactly vanish. Equivalently, any such non-vanishing configuration is projected out by the path integral, resulting in a restriction to flat connections. This holds also both for the intermediate reduction (as realized by the Lagrange multiplier \eqref{eq:FlatnessForce}) as well as for consistent truncations and allows for the localization of the flux onto defects. Away from this limit, when $X_5$ is of finite volume, kinetic terms contribute to the equations of motion and imply the usual non-localized flux profiles associated with electromagnetic sources.
In contrast, the KK 5-brane boundaries of the fluxbrane remain localized when pushed into the physical AdS bulk and source this bulk flux profile. Recall, in the top-down 10D background these are locally simply codimension-four metric singularities. Similar comments hold regarding the singularities associated with codimension-4 fixed loci upon replacing codimension-four metric singularities with codimension-6 singularities.

Supersymmetric fluxbranes have been studied in AdS spaces of various spacetime dimensions. For example, in \cite{Lu:2003iv} smooth configurations are studied, see also \cite{Arav:2024exg} for related discussion on monodromy defects, while considerations in \cite{Bomans:2024vii} include properly quantized fluxbranes sourced by KK 5-branes realized as local patches modeled on $\mathbb{R}^4/\mathbb{Z}_k$. The latter are derived from the bubbling solutions of \cite{Liu:2007rv, Chong:2004ce}.

To proceed further we now focus on the heavy monodromy defects constructed in \cite{Arav:2024exg}
and deform / boost the configurations to construct the bulk duals of symmetry operators for $G = \mathrm{Isom}(X)$.

\subsection{Fluxbranes in the Bulk} \label{sec:Bulk}

So far we have focused on topological aspects of the fluxbrane configurations used to build our symmetry operators. We now proceed to build the corresponding gravity dual defects. We begin by focusing on the special case of fluxbrane defects in AdS$_5 \times S^5$ and then explain how these considerations generalize to more general backgrounds of the form AdS$_5 \times X_5$. We use this to read off the tension of the corresponding defects, i.e., we confirm the expectations from the argument in \cite{Heckman:2024oot} that these defects have a tension and thus couple to local fluctuations of the bulk metric. As such they are best viewed as dynamical objects in the bulk.

Quite auspiciously, many aspects of the relevant defect configurations for AdS$_5 \times S^5$
have recently been worked out in \cite{Arav:2024exg, Bomans:2024vii}. As such, it is essentially enough to reinterpret
these results to extract the relevant physical data for our symmetry operators. We focus on the case of defects parameterizing the Cartan
$U(1)^3 \subset \mathrm{Isom}(S^5)$. In this case, we have three distinct gauge fields to pay attention to, including their asymptotic values. The case AdS$_5 \times X_5$ then follows as a further (mild) generalization of these considerations.

As a side note, in general, the preferred consistent truncation depends on the space $X_5$, and even given a fixed $X_5$ there are distinct consistent truncations describing different sectors of solutions in the 10D uplift, see for example \cite{Liu:2007rv} and \cite{Khavaev:2000gb, Bobev:2010de}. Consequently, in describing both defect and symmetry operators associated with some isometry subgroup through an uplift of a consistent truncation, one should not expect to recover all possible such operators from any one truncation. This restriction extends to studying configurations of these defects, for example, fusions and intersections of arbitrary defects are generally not accessible given any one truncation.

With these shortcomings remarked we begin by first focussing on a single heavy defect. One feature of interest is the differences between various field profiles at the conformal boundary and in the AdS bulk. These differences inform the fate of a symmetry operator fluxbrane when it is pushed from the SymTFT\,/\,SymTh sliver into the AdS bulk.

\subsubsection{4D $\mathcal{N}=4$ SYM}

We begin by reviewing some features of the 5D solutions in \cite{Arav:2024exg} and \cite{Bomans:2024vii}, and their 10D uplifts.
To match more directly with the discussion in \cite{Arav:2024exg, Bomans:2024vii} in this subsection we adopt the conventions presented there (as opposed to the more natural topological conventions used in section \ref{sec:DEFECTS}). We consider the consistent truncation to a $U(1)^3$ gauge theory associated with the Cartan of $\mathrm{Spin}(6)$. The bulk action is:
\begin{equation}
S_{\mathrm{bulk}} = \frac{1}{16 \pi G_{5}} \int \dd^{5} x \sqrt{g} \mathcal{L}_{\mathrm{bulk}},
\end{equation}
where the 5D Newton's constant is related to the AdS$_5$ radius via:
\begin{equation}
\frac{L^2}{16 \pi G_5} = \frac{N^2}{8 \pi^2}.
\end{equation}
In a mostly minus\footnote{We stress that this is to adhere to the conventions already given in \cite{Arav:2024exg}.} metric sign convention
the bulk Lagrangian density is:
\begin{equation}
\mathcal{L}_{\mathrm{bulk}} = -\frac{1}{4} \mathcal{R}
- \frac{1}{4} (e^{4 \beta_1 - 4 \beta_2} F^{(1)}_{\mu \nu} F^{(1) \mu \nu} + e^{4 \beta_1 + 4 \beta_2} F^{(2)}_{\mu \nu} F^{(2) \mu \nu}
+ e^{-8 \beta_1} F^{(3)}_{\mu \nu} F^{(3) \mu \nu})+ ...,
\end{equation}
where the ``...'' includes 5D Chern-Simons terms for the gauge fields as well as
the kinetic terms and effective potentials for the scalars $\beta_1,\beta_2$.
In these units, the gauge covariant derivative appears via:
\begin{equation}
D_{A} = \dd + g A,
\end{equation}
and in our particular case we have the scaling relation $g = 2 / L$.

As found in \cite{Arav:2024exg, Bomans:2024vii}, the flux defect solutions of interest to us are of the form:
\begin{equation}
\label{eq:Sol}
\begin{aligned}
    \dd s^2&=f(r) \dd s^2_{\text{AdS}_3} + g(r) \dd r^2 +h(r)\dd\phi^2\,,\\[0.25em]
    A_1^{(I)}&= a_0^{(I)}(r) \dd\phi\,,
\end{aligned}
\end{equation}
with further equations specifying the remaining field content (scalars). References \cite{Arav:2024exg} and \cite{Bomans:2024vii} make some different choices for scalar profiles, but these distinctions are immaterial for our primary focus, which is the flux defect. The flux defect solutions describe a monodromy defect filling an AdS$_3$ slice. The coordinates $(r,\phi)$ are polar coordinates for the remaining two transverse dimensions and the defect sits at $r=0$. This follows from the local coefficient functions $a_0^{(I)}(r) $ taking the form
\begin{equation}
\label{eq:connectioncomponent}
    a_0^{(I)}(r) =\mu^{(I)}+\dots+\frac{L^2 j_0^{(I)}}{r^2}+\dots\,.
\end{equation}
The functions $f,g,h$ are asymptotically such that \eqref{eq:Sol} is asymptotically AdS$_5$ with radius $L$. The constants $\mu^{(I)}$ set an asymptotically flat background for the connections $A_1^{(I)}$. The constants $j^{(I)}_0$ set the background values of the dual $U(1)^{(I)}$ field theory current components $J^{(I)}$ and we turn them off. In terms of the Lie algebra data for the various flux defects we further have:
\begin{equation}
\alpha^{(I)} = g \mu^{(I)}.
\end{equation}

Next, we remark that this solution is fibered by copies of AdS$_3\times S^1$ labeled by $r$ and the functions $f,g,h$ are such that asymptotically $r\rightarrow \infty$ the space resembles AdS$_3\times S^1$ in an ambient AdS$_5$. In this limit, one has the boundary metric
\begin{equation}
    \dd s^2 \propto \frac{1}{\rho^2} \left( \dd t^2-\dd x^2 - \dd\rho^2 -n^2\rho^2 \dd\phi^2 \right)\,,
\end{equation}
in Poincar\'e coordinates with $(t,x,\rho)$ parameterizing the AdS$_3$.

Now, consider a Weyl rescaling which takes this asymptotic metric to that of $\mathbb{R}^{3,1}$ in such a way that the boundary of AdS$_3$ is mapped onto $\mathbb{R}^{1,1}\subset \mathbb{R}^{3,1}$. The metric on $\mathbb{R}^{3,1}$ is flat up to a codimension-2 conical deficit, determined by $n>0$, centered on $\mathbb{R}^{1,1}$.
Concretely, the conical deficit is $\chi = 2 \pi(1-n)$. Whenever $\phi$ is normalized to have period $2\pi$ then $n=1$ corresponds to no deficit angle. It is in this Weyl frame that at $\rho=0$ in the CFT dual one has a heavy monodromy defect.

The conical deficit $n$ is related to the radius of the asymptotic AdS$_5$, which sets the 5D gauge coupling $L=2/g$. Here the three gauge couplings of $U(1)^3$ agree and are all equal to $g$. The conical deficit is then determined from the asymptotic flat gauge field profiles, which determines how much magnetic flux is localized to the defect, following the relation
\begin{equation}
    \alpha^{(1)} + \alpha^{(2)} + \alpha^{(3)} = (1-n) \kappa\,,
\end{equation}
where $\kappa$ is a dimensionless number.

Here we have given the ``main branch'' solution of \cite{Arav:2024exg} which is continuously connected to the case $n=1$ and which is also used in \cite{Bomans:2024vii}. The defects constructed there preserve, in two dimensions either $\mathcal{N}=(0,2)$ supersymmetry (when $\kappa = +1$) or and $\mathcal{N}=(2,0)$ supersymmetry (when $\kappa = -1$).

By the above, in the asymptotic limit, we are thus entitled to speak of a ``fluxbrane'' with tension \eqref{eq:Tension}. For completeness, we combine this as the general formula:\footnote{Recall that in this expression all the $\alpha^{(i)}$ take values in the interval $[0, 1)$, and in order to self-consistently neglect backreaction effects we require all the $\alpha^{(i)}$ be close to zero.}
\begin{equation}\label{eq:tensioncomputed}
T_{\ast}^{\mathrm{AdS-Defect}} = \frac{\chi}{8 \pi G_{5}} = \frac{2 \pi (\alpha^{(1)} + \alpha^{(2)} +\alpha^{(3)})}{8 \pi G_{5}}.
\end{equation}
Deeper in the bulk, when the suppressed terms in \eqref{eq:connectioncomponent} become relevant, the flux spreads out and is no longer localized to codimension-2, i.e., the thickness of the brane grows, ultimately becoming comparable to the length scale $L$ of the bulk gravitational background.

One crucial feature of the solutions of \cite{Bomans:2024vii}, which include the case in which the $X_5/U(1)$ is a manifold with boundary, i.e., the fluxbrane is realized together with a KK 5-brane source (corresponding to the boundary), is that, while the abelian flux delocalizes in the gravitational bulk, the KK 5-brane, in contrast, remains as an overall localized codimension-4 singularity in the bulk.

Finally, we consider the symmetry operators. From our original discussion in section \ref{sec:BOOST}, we can consider two of the discussed monodromy defects of equal but opposite monodromy, and individually deform their bulk support and reconnect these. In the above discussion, after Weyl transformations, these supports would be two 3D half-spaces in the bulk with asymptotic boundary $\mathbb{R}^{1,1}$ of opposite orientation. These glue to $\mathbb{R}^{2,1}$ at finite bulk radius.

Such solutions have not been constructed to our knowledge, and instead, we opt to discuss a slightly different realization of a symmetry operator as related to a single heavy defect (rather than a pair). If we broaden our considerations and allow the 4D spacetime manifold to be AdS$_3\times S^1$ then, before the Weyl transformation, the above solutions already realize a topological symmetry operator for a metric isometry via a bulk fluxbrane. There, the defect core remains located at $r=0$ which is now away from the boundary. Flat holonomy along the $S^1$ remains, and we can localize it via a gauge transformation to a single transition function at a point on the $S^1$. This transition function realizes a topological symmetry operator of 4D $\mathcal{N}=4$ Super-Yang-Mills (with spacetime AdS$_3\times S^1$) sitting at a point on $S^1$ and filling AdS$_3$.

Summarizing, we have now lifted the symmetry operator to an explicit 10D metric profile which is topologically a fibration (with a singular sublocus) of $S^5$ over AdS$_5$.

\subsubsection{AdS$_5 \times X_5$  Backgrounds}

With these facts established in the example of 4D $\mathcal{N}=4$ Super-Yang-Mills theory, we now move to discuss the general case of a stack of $N$ D3-branes probing a local Calabi-Yau cone $Y = \mathrm{Cone}(X_5)$ with $X_5$ a Sasaki-Einstein five-manifold. Explicit examples of such backgrounds include toric Calabi-Yau threefolds (see e.g., \cite{Gauntlett:2004hh, Martelli:2004wu, Cvetic:2005ft}). In this case, the worldvolume of the D3-branes produces a 4D $\mathcal{N} = 1 $ SCFT and the gravity dual is of the general form AdS$_5 \times X_5$. The R-symmetry is dual to a particular $U(1) \subset \mathrm{Isom}(X_5)$ but in principle there can be several different $U(1)$'s. Determining the precise linear combination corresponding to this IR R-symmetry is typically realized by a-maximization \cite{Intriligator:2003jj} / Z-minimization \cite{Martelli:2005tp}. A broad class of examples can be obtained via orbifolds of the form $Y = \mathbb{C}^{3} / \Gamma$ for $\Gamma$ a finite subgroup of $SU(3)$. Observe that the commutant of $\Gamma$ in $\mathrm{Spin}(6)$ results in a natural class of isometries. As such, we can simply inherit the same gauge field configurations used in the case of 4D $\mathcal{N} = 4$ SYM. In particular, formulae such as the tension formula of equation (\ref{eq:tensioncomputed}) still apply; we simply need to specify the asymptotic profile for the bulk gauge fields near the core of the flux defect.

\section{TFT of the Symmetry Operator} \label{sec:PROPERTIES}

In previous sections, we presented a general method of relating singular geometric fibrations to symmetry operators.
Our plan in this section will be to determine some properties of the relevant topological field theory supported on this symmetry operator. These topological field theories constructed via fluxbranes are only localized in codimension-2 in the SymTFT\,/\,SymTh sliver. In the gravitational bulk the flux spreads out, and the topological field theory smears across the full bulk.

To extract the topological terms on the worldvolume of our fluxbrane singularity, we shall make use of its corresponding SymTFT$_{\mathrm{defect}}$.
We emphasize that this is \textit{not} the SymTFT$_{D+1}$ associated with a topological subsector
of our holographic model, rather, it is the SymTFT of the defect itself.

There is by now a well-defined prescription for reading off the contributions to the SymTFT for QFTs engineered via string theory.
See, e.g., references \cite{Cvetic:2024dzu, Apruzzi:2021nmk, Baume:2023kkf, GarciaEtxebarria:2024fuk, GarciaEtxebarria:2024jfv, Gagliano:2024off, Yu:2024jtk, Tian:2024dgl, Najjar:2025rgt} for details of this construction.
The main idea is that one begins with a conical geometry of the form $\text{Cone}(X) = Y$.
Starting from the higher-dimensional gravitational background, we then perform a formal dimensional reduction along the directions $X$.
For example, in the case of a geometry such as $\mathbb{C}^{n} / \Gamma$, this involves reduction along the generalized lens space
$S^{2n-1} / \Gamma$.
This procedure also works in situations where the singularities of the conical geometry extend to the boundary $\partial Y = X$.
In such situations, one considers a filtration to a nested collection of relative symmetry theories, as in reference \cite{Cvetic:2024dzu}. Importantly, one can also consider situations in which the original QFT has been deformed into a tree-like structure.
In this case, the SymTFTs form a ``SymTree'' joined by a non-topological junction \cite{Baume:2023kkf}.

We start with the simplest situation in which we have an isolated singularity at the tip of the cone $\text{Cone}(X) = Y$.
We label the bulk action for the corresponding SymTFT as $S_{\text{blk}} \left[  \left\{ \Phi \right\} \right]$, where $\Phi$ is shorthand for all of the bulk gauge fields present in the system.
Suppose next that we have bulk $p$-form gauge field $\Phi_{p}$.
Given this potential, there is a corresponding $p$-dimensional object that couples to this potential.
Let us assume that this $p$-dimensional object can serve as a genuine heavy defect.\footnote{This depends on consistent (i.e., non-anomalous) choices for the topological boundary conditions of the SymTFT.}
In the resulting QFT, there is a corresponding $(p-1)$-dimensional object which is the boundary for the heavy defect.

Suppose next we switch on a background value of $\Phi_{p}$ in the worldvolume of the boundary QFT.
Observe that this induces a source for the objects which couple to the potential.
As such we can induce a background flux for dynamical states of the theory.
Denote by $\phi_{p-1}$ the corresponding $(p-1)$-form potential which couples to these states (such couplings can take the form of Chern-Simons-like terms).
We can then read off a corresponding topological term directly in the QFT:
\begin{equation}
    S_{\text{top}} \supset \phi_{p-1} \wedge   \frac{\delta S_{\text{blk}}}{\delta \Phi_{p}} \bigg \vert_{\partial} \, ,
\end{equation}
where the notation $|_{\partial}$ indicates to evaluate all bulk fields (treated as background fields) on the {topological} boundary.
More formally, one views the (relative) QFT as embedded in the SymTFT and takes the pullback of all bulk fields onto the boundary.

One can perform some basic checks that this reproduces known topological terms.
For example, in the case of a D-brane, this leads to the expected lower-degree WZ terms, essentially because these are produced from ``branes ending in branes''.
Additionally, observe that in Yang-Mills theory, this leads to a BF-type topological coupling which can be viewed as coupling a QFT to a TQFT,
essentially switching the global form of the gauge theory
\cite{Kapustin:2014gua}.

This is almost the full answer, but to complete the story we also need to include the ``brane charge'' of the QFT itself.
Since these objects arise as KK monopoles, we can immediately add in ``by hand'' the relevant top degree charge.
More formally, we return to section \ref{ssec:REDUCTION} and integrate over the magnetic dual gauge field $A_{d}$, where $d$ denotes the
dimension filled by the defect QFT.
Taking this into account, we reach our proposed form for the defect topological terms:
\begin{equation}
    S_{\text{top}} = \alpha \int A_{d} + \int \phi_{p-1} \wedge  \frac{\delta S_{\text{blk}}}{\delta\Phi_{p}} \bigg\vert_{\partial} \, .
\end{equation}
for some constant $\alpha$.

\begin{figure}
\centering
\scalebox{0.8}{
\begin{tikzpicture}
	\begin{pgfonlayer}{nodelayer}
		\node [style=none] (0) at (-2, 1.5) {};
		\node [style=none] (1) at (-2, -1.5) {};
		\node [style=none] (2) at (2, 1.5) {};
		\node [style=none] (3) at (2, -1.5) {};
		\node [style=Star] (4) at (0, 0) {};
		\node [style=none] (5) at (-2, 0) {};
		\node [style=none] (6) at (-3, -1) {};
		\node [style=none] (7) at (-3, 1) {};
		\node [style=none] (8) at (-1, -2.375) {};
		\node [style=none] (9) at (1, -2.375) {};
		\node [style=none] (10) at (-3.5, 0) {$x_\perp$};
		\node [style=none] (11) at (0, -2.75) {$r$};
		\node [style=none] (12) at (-2, 2) {};
		\node [style=none] (13) at (2, 2) {};
		\node [style=none] (14) at (2, -2) {};
		\node [style=none] (15) at (-2, -2) {};
		\node [style=none] (16) at (0, 0.75) {$\Sigma_3$};
		\node [style=none] (17) at (4, 0) {};
		\node [style=none] (18) at (0, 2.625) {SymTFT / SymTh};
		\node [style=NodeCross] (19) at (-2, 0) {};
		\node [style=none] (20) at (3.5, 0.25) {Topological};
		\node [style=none] (21) at (3.5, -0.25) {Boundary};
        \node [style=none] (22) at (0, -1.5) {};
		\node [style=none] (23) at (0, -1.9) {};
	\end{pgfonlayer}
	\begin{pgfonlayer}{edgelayer}
		\draw [style=ThickLine] (0.center) to (1.center);
		\draw [style=ThickLine] (2.center) to (3.center);
		\draw [style=ArrowLineRight] (8.center) to (9.center);
		\draw [style=ArrowLineRight] (6.center) to (7.center);
		\draw [style=DashedLine] (4) to (5.center);
		\draw [style=DottedLine] (12.center) to (0.center);
		\draw [style=DottedLine] (13.center) to (2.center);
		\draw [style=DottedLine] (14.center) to (3.center);
		\draw [style=DottedLine] (15.center) to (1.center);
        \draw [style=BlueLine] (4) to (22.center);
		\draw [style=DottedBlueLine] (22.center) to (23.center);
	\end{pgfonlayer}
\end{tikzpicture}}
\caption{We consider the theory on $\Sigma_3$ as a QFT in its own right. Then we can consider heavy defects of this theory as engineered by objects extending in the additional ambient dimensions which terminate on the $\Sigma_3$ (blue line). However, when these extend at fixed AdS radius $r$ these defects are interpreted as symmetry operators in the dual CFT, and overall we construct a configuration of symmetry operators.}
\label{fig:DualityDefectLike2}
\end{figure}
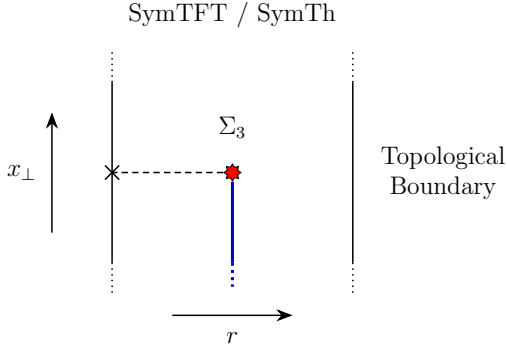

There is a complementary geometric perspective that sheds more light on how the worldvolume theory of the symmetry operator interacts with the ambient field theory. For this, instead of considering the full SymTFT of the worldvolume theory we focus on its defects. Certain configurations of such defects may be interpreted as symmetry operators of the ambient field theory. See figure \ref{fig:DualityDefectLike2}.

To make this precise, consider any defect (such as an isometry defect) that deforms the direct product $M\times X$ to include an exceptional fiber type $X'$. Here $M$ are the ``external'' dimensions, for example, $M=\text{AdS}_{D+1}$, but flat spacetimes are generally also permitted. We can push the generic fiber $X$ on top of this defect realizing via this deformation a mapping $X\rightarrow X'$ with homology\footnote{Here we are considering standard integral singular homology groups. Depending on the singular structures of $X$ and various fluxes, twisted Chen-Ruan orbifold cohomology groups \cite{Chen:2000cy} would be the appropriate generalization, although we will not need this machinery here.} lift
\be\label{eq:collapseeq}
\pi_n\,: ~H_n(X)\rightarrow H_n(X')\,.
\ee
Then, consider any $p$-dimensional family of elements in Ker$(\pi_n)$ and construct non-compact cycles of dimension $p+n$. Wrapping a brane on this constructs a defect for the theory supported along the degenerate fiber. However, from the perspective of the QFT this constructs a configuration of SymTFT\,/\,SymTh operators, if the $p$-dimensional family is at a constant AdS radius. The operator constructed by brane wrapping terminates on the operator associated with topology change, due to the respective homology class of $X$ trivializing as an element in $H_n(X')$. It follows there exists a chain of degree $n+1$ which closes off the family of $n$-cycles such that the overall family can no longer be deformed away from $X'$. See also \cite{Lawrie:2023tdz} for a related discussion on such fiber degenerations.

The upshot of this more narrow perspective is that the symmetry operators of the ambient field theory are under better control. Various QFTs are realized in string constructions only after certain decoupling limits, and understanding which brane wrappings at infinity remain as topological operators acting on the QFT is non-trivial. However, once these are determined, by the above arguments we can check via line \eqref{eq:collapseeq} if their wrapping locus collapses at a metric defect, and whether it is consequently endable on this defect. The world volume theory on the original symmetry operator then couples to the background fields associated with the ambient field theory symmetry operators constructed from the kernel of line \eqref{eq:collapseeq}.

\subsection{Illustrative Example: 4D $\mathcal{N}=4$ SYM}

Let us return to the example of 4D $\mathcal{N}=4$ SYM as constructed from a stack of $N$ D3-branes probing $\text{Cone}(S^5)=\mathbb{C}^3$ with gravity dual given by IIB on AdS$_5\times S^5$ with a 5-form flux. In the asymptotic SymTFT sliver, the topological operator associated with the 0-form isometry is a codimension-2 fluxbrane geometrically characterized by monodromy of the $S^5$ along linking paths and degenerate exceptional fibers projecting to the fluxbrane locus.

To determine properties of the corresponding SymTFT for our KK 5-brane and flux configuration, we shall find it useful to proceed via a dual characterization of the KK 5-branes in M-theory. Along these lines, we first observe that for special choices of metric isometry and specific values of the flux defect parameter $\alpha \in \mathbb{Z}$ (i.e., the ``trivial'' case) we actually wind up with a well-known dual. Along these lines, consider again our stack of D3-branes filling the first factor of the 10D spacetime $(\mathbb{R}^{1,1} \times \mathbb{C}_u) \times (\mathbb{C}_z \times \mathbb{C}^2)$. A supersymmetric quotient singularity of the form $(\mathbb{C}_u \times \mathbb{C}_{z}) / \mathbb{Z}_k$ introduces a particular 2D supersymmetric defect into the 4D $\mathcal{N} = 4$ SYM of the same sort considered in \cite{Bomans:2024vii}.\footnote{When the deficit angle and holonomies are quantized and the deficit angle takes value $n=1  /k$, for some integer $k$, it was shown in \cite{Bomans:2024vii} that one has an $U(1)^3$ orbibundle locally modeled on the global quotient $(\mathbb{C}\times U(1)^3) / \mathbb{Z}_K$ where the identifications are:
\begin{equation}
    \left( re^{i \phi}, e^{ i \theta_1}, e^{i \theta_2}, e^{ i \theta_2} \right) \sim \left( re^{ i \phi +  i/k}, e^{ i \theta_1+  i m_1/k}, e^{ i \theta_2+  i m_2/k}, e^{ i \theta_2 + i m_3/k} \right) \,,
\end{equation}
with  $m_i\in 2\pi \mathbb{Z}$ specifying the holonomies.

It is worth noting here that in these sorts of orbifold constructions (and their deformations to capture more general flux profiles), one ought to expect differences in the sourced stress energy. At the level of the topological terms, however, these distinctions do not really matter. In particular, so long as we have the same monodromy structure from the accompanying $X_5$ fibration over AdS$_5$ and its restriction to the SymTFT$_{D+1}$ sliver, we can still read off the relevant topological terms.}

The monodromy generated by this case is $\alpha = k$, i.e., this implements trivial linking with the KK momenta. That being said, this is a well-known object in type IIB backgrounds: it implements a 6D $\mathcal{N} = (2,0)$ SCFT of type $A_{k - 1}$. Our interest is in deformations of this sort of configuration so that we actually have non-trivial linking.

It is here that the M-theory characterization will prove to be quite useful. Along these lines, observe that precisely the same sort of 6D SCFT is implemented by a stack of $k$ coincident M5-branes. In this phrasing, KK 5-branes are mapped onto M5-branes. M5-branes are codimension-5 and linked by 4-spheres. The dual of the KK fluxbrane is codimension-4 and linked by 3-spheres. So, to figure out properties of the SymTFT$_{\mathrm{defect}}$ for the flux defect, it suffices to consider M-theory backgrounds with an asymptotic 3-form potential specified by more general $\alpha$:
\begin{equation}\label{eq:C3period}
    2 \pi \alpha=\int_{S^3}C_3\,.
\end{equation}

Let us argue for this same linking condition directly by tracking how string dualities behave in the SymTFT$_{D+1}$ sliver. In this region, we can neglect metric data and focus exclusively on topological fibration structures. With this in mind, observe that the monodromy defect induces a circle fibration over a sliver of AdS$_{D+1}$. T-dualizing this, we get in type IIA NS-6 fluxbranes with period $B_2$ on a linking $S^2$. Lifting this to M-theory results in equation (\ref{eq:C3period}).

Therefore, starting from the topological terms of 11D supergravity
\begin{equation}
    S_{\text{SUGRA}}^{\text{top}} = \frac{2\pi }{6}\int \frac{C_3}{2\pi} \wedge \left( \frac{G_4}{2\pi} \wedge \frac{G_4}{2\pi} +\frac{p_1\wedge p_1-4p_2}{32} \right) \, ,
\end{equation}
with $G_4=dC_3$ and $p_i$ the $i$-th Pontryagin classes, we find the SymTFT of the fluxbrane dual to be
\begin{equation}
    S_{\text{blk}}=\frac{2\pi \alpha}{2}\int  \left(  \frac{G_4}{2\pi} \wedge \frac{G_4}{2\pi} +\frac{p_1\wedge p_1-4p_2}{96} \right) \,.
\end{equation}
By our general discussion, this now results in
\begin{equation}
\label{eq:StopC2ZK2}
    S_{7D}^{\text{top}} = \alpha \int F_{7} + \frac{\alpha}{4\pi} \int\dd b_{2}^{(-)} \wedge \dd  C_{3} \,,
\end{equation}
where we have normalized $F_7$ to have integral quantized flux and dropped the metric terms which do not contribute for integrals supported on $\Sigma_3\times B_4$. The fluxbrane worldvolume gauge field $b_{2}^{(-)}$ couples to string like defects and its field strength $h_{3}^{(-)}=\dd b_{2}^{(-)} $ is anti-self-dual, similar to the gauge field living on an M5-brane. The reason for this identification is evident in settings where the fluxbrane is sourced by KK 5-branes. In the topological sliver at the conformal boundary, the fluxbrane is characterized by the same circle collapsing at the KK 5-brane locus. Correspondingly, the extended objects, constructed by wrapping this cycle, and which can end on the KK 5-brane, can also end on a fluxbrane. The gauge field of the fluxbrane therefore must restrict to that on the KK 5-brane which in the M-theory dual is the well-known gauge 2-form with anti-self dual field strength.

Compactifying further on $B_4=X_5/U(1)$ the topological terms then descend to
\begin{equation}
\label{eq:SymOpWithbeta}
    T_{3D}^{\text{top}} \left[  \alpha, \beta \right]  = \alpha \int A_{3} + \frac{\alpha\beta}{2} \int h_{3}^{(-)} \, .
\end{equation}
Here, $\alpha \in [0,1)$ and $\beta$ is determined by the background as the integral of $\dd C_3$.

In IIB, the 3-form potential $C_3$ is given by the RR 4-form potential $C_4^{\text{RR}}$ integrated over the circle of the intermediate reduction. However, the degrees of freedom of the 4D worldvolume theory of the stack of D3-branes arise exclusively from open strings that do not couple to $C_4^{\text{RR}}$. The symmetry operator acting on the 4D theory is therefore simply
\begin{equation}\label{eq:SymOpN4}
    T_{3D}^{\text{top}} \left[  \alpha \right]  = \alpha \int A_{3}\, ,
\end{equation}
describing an invertible symmetry. Indeed, the R-symmetry of 4D $\mathcal{N}=4$ SYM is an invertible 0-form symmetry. We note that \eqref{eq:SymOpN4} is topological operator for R-symmetry rotations contained in the $U(1)$ subgroup of the full R-symmetry group as specified by $B_4=X_5/U(1)$.

\subsection{A More Involved Example: D3-Brane Probes of $\mathbb{C}^3/\mathbb{Z}_3$}
\label{sec:moreinvolved}

We consider a stack of $N$ D3-branes probing the Calabi-Yau orbifold $\mathbb{C}^3/\mathbb{Z}_3$ with weights $(1,1,1)$. The gravity dual of this setup is AdS$_5\times S^5/\mathbb{Z}_3$ and the R-symmetry group is geometrically specified by a $U(1)$ which lifts to the canonical $S^1/\mathbb{Z}_3 \rightarrow S^5 /\mathbb{Z}_3 \rightarrow\mathbb{CP}^2$ Hopf fibration on the covering space.

Unlike the previous example, the associated 4D quiver gauge theory exhibits a discrete $\mathbb{Z}_3$ worth of 0-form and 2-form defects and symmetries \cite{Gukov:1998kn, Heckman:2022xgu} that also act on the field content. These defects are constructed by wrapping D3-branes on non-compact 4- and 2-cycles respectively. The previous construction for R-symmetry monodromy defects goes through until line \eqref{eq:SymOpWithbeta}, which we repeat for convenience here:
\begin{equation}\label{eq:SymOpWithbeta2}
    T_{3D}^{\text{top}} \left[  \alpha, \beta \right]  = \alpha \int A_{3} + \frac{\alpha\beta}{2} \int h_{3}^{(-)} \, .
\end{equation}
However, now, we can not drop the last term due to the additional defects and symmetry operators constructed from D3-branes. In particular, taking the field strength $h_3^{(-)}$ at face value, we encounter topological $1$-brane and $(-1)$-brane defects localized in the 3D symmetry operator (which of these depends on the topological boundary conditions). To make progress, we first establish that $\beta$ is indeed non-vanishing. Then we construct these $1$-branes / $(-1)$-branes as endpoints of the operators acting on the 0-form and 2-form defects.

The constant $\beta$ is given by the integral of $\dd C_3$ over $B_4$. As $B_4=\mathbb{CP}^2=S^5/U(1)$ with $\mathbb{Z}_3\subset U(1)$, we parameterize the $S^5/\mathbb{Z}_3$ as
\begin{equation}
\label{eq:Fibration2}
    S^1/\mathbb{Z}_3~\hookrightarrow~ S^5/\mathbb{Z}_3 ~\rightarrow~ \mathbb{CP}^2\,.
\end{equation}
We previously also identified ${C}_3$ as the IIB RR gauge potential $C_4^{\text{RR}}$ integrated over the circle which collapses along $ \mathbb{CP}^2$. Consider therefore the flux integral
\begin{equation}
    N=\int_{S^5/\mathbb{Z}_3} \frac{F_5^{\text{RR}}}{2\pi}= \int_{\mathbb{CP}^2} \int_{S^1/\mathbb{Z}_3} \frac{F_5^{\text{RR}}}{2\pi}= \int_{\mathbb{CP}^2}\frac{\dd C_3}{2\pi} \, ,
\end{equation}
where $\int_{S^1/\mathbb{Z}_3}$ is the fiber integration as given by the Gysin homomorphism which features in the Gysin sequence as associated to the circle bundle \eqref{eq:Fibration2}. Consequently, we have $\beta=N$. For generic $\alpha$, the second term in \eqref{eq:SymOpWithbeta2} is therefore present.

This is completely expected upon noting that the topological operators acting on 0-form and 2-form defects are constructed from D3-branes wrapped on the generator of $H_1(S^5/\mathbb{Z}_{3})$ and  $H_3(S^5/\mathbb{Z}_{3})$ asymptotically in the boundary. Any representative of these classes involves the circle which is collapsed along $B_4$. More precisely, the homology class of this KK circle is precisely a generator of  $H_1(S^5/\mathbb{Z}_{3})$ and any representative of a generator of $H_3(S^5/\mathbb{Z}_{3})$ can be presented as an $S^3/\mathbb{Z}_3$ whose Hopf fiber on its own also gives a class which generates $H_1(S^5/\mathbb{Z}_{3})$. As such the generators of $H_1(S^5/\mathbb{Z}_{3})$ and  $H_3(S^5/\mathbb{Z}_{3})$, when pushed into $\mathbb{CP}^2$ collapse to a point and the hyperplane class $\mathbb{CP}^1$ respectively.

Asymptotically, we therefore have new wrapping loci to construct symmetry operators following the general ideas of \cite{Heckman:2022muc}. To see this, consider AdS$_5\times {S}^5/\mathbb{Z}_3$ deformed by the insertion of an R-symmetry defect in the asymptotic sliver. We can now build a non-compact 4-cycle by fibering $S^1/\mathbb{Z}_3$ over a 3-surface in AdS$_5$ that terminates at the R-symmetry operator. Similarly, we obtain a non-compact 4-cycle by fibering $S^3/\mathbb{Z}_3$ over a line in AdS$_5$ that also necessarily terminates at the R-symmetry operator. The endpoints of the 3-surface and the line on the worldvolume of the R-symmetry operator are respectively a 2-surface and a point. Neither of these 4-cycles can be deformed away from the R-symmetry defect insertion. We now wrap D3-branes on these non-compact 4-cycles. When realizing symmetry operators in the dual field theory we have therefore constructed 0-form and 2-form symmetry operators which end on the R-symmetry operator, and which of the former is present depends on the topological boundary condition fixing the global form of the 4D field theory (see figure \ref{fig:DualityDefectLike2}).

Let us now consider the above defects, denoted $T_\alpha$, in more detail. We have
\begin{equation}
\begin{aligned}
    \mathcal{U}_\alpha&=\exp \left( i \alpha \int_{\Sigma_3} A_{3}  \right) \,,\\[0.2em]
    \mathcal{C}_\alpha&=\frac{1}{|H_3(\Sigma_3;\mathbb{Z})|}\int Db_2^{(-)}\exp \left(  i \, \frac{\alpha N}{2} \int_{\Sigma_3} h_{3}^{(-)}  \right) \,,\\[0.65em]
    T_\alpha&=\mathcal{U}_{\alpha\,}\mathcal{C}_\alpha\,,
\end{aligned}
\end{equation}
where $\int Db_2^{(-)}$ is the path integral over the anti-self-dual 2-form $b_2^{(-)}$ with field strength $h_3^{(-)}$. Then, $\mathcal{U}_\alpha$ describes the invertible U(1) R-symmetry of the 4D $\mathcal{N}=1$ SCFT. 

In contrast, $\mathcal{C}_\alpha$ is a condensation operator which realizes a projection for generic values of $\alpha$. Indeed, the field strength $h_3^{(-)}$ is associated via a 3-form potential $C_3$ obtained from reduction of $C_{4}^{RR}$ on a circle. In the backgrounds under consideration, however, there is no field configuration in the D3-brane worldvolume theory which activates this $C_3$. As such $\mathcal{C}_\alpha$ does not act on any local operators, confirming the general expectation that these R-symmetry topological operators act invertibly on local operators. That being said, we have also seen that there are 0-form and 2-form symmetry operators constructed from D3-branes in this setting (which one is realized depends on the overall polarization, i.e., choice of topological boundary conditions). We have also argued geometrically that these can end on the isometry defect $T_\alpha$ (see line \eqref{eq:collapseeq}). In 4D, the avatar of this geometric reasoning is precisely the operator $\mathcal{C}_\alpha$. When the 4D 0-form or its dual 2-form operators extend into $T_\alpha$ the path integral in $\mathcal{C}_\alpha$ induces Dirchlet boundary conditions, allowing these operators to terminate. Denoting 0-form and 2-form symmetry operators by $\mathcal{N}^{(0)},\mathcal{N}^{(2)}$, we have the fusion relations
\be
\mathcal{C}_{\alpha\,} \mathcal{N}^{(0)}=\mathcal{C}_{\alpha\,} \mathcal{N}^{(2)}=\mathcal{C}_\alpha \,.
\ee
Further, noting that $\mathcal{C}_\alpha$'s only role in the field theory is to supply Dirichlet boundary conditions for $\mathcal{N}^{(0)},\mathcal{N}^{(2)}$, we see that it is independent of the value of generic $\alpha\neq 0$. We therefore simply denote it by $\mathcal{C}$. Further, we have the fusion rules
\begin{equation}
\begin{aligned}
    \mathcal{U}(\alpha)\mathcal{U}(\alpha')&=\mathcal{U}(\alpha+\alpha') \,, \\
    \mathcal{C}\:\!\mathcal{C}&=\mathcal{C} \,, \\
\end{aligned}
\end{equation}
which in particular determine the self-fusion of $T_\alpha$. We emphasize again, that $T_\alpha$ realizes an invertible continuous 0-form symmetry on local operators of the 4D SCFT carrying charge under the isometry.

\section{Non-Holographic Example}\label{sec:NONHOLO}

Our primary focus in this work has been on holographic examples involving the isometries of $X$ in backgrounds of the form AdS$_{D+1} \times X$. On the other hand, these backgrounds arise from the near horizon limit of branes probing special holonomy cones of the form
$Y = \mathrm{Cone}(X)$. On general grounds, we expect that isometries / diffeomorphisms can be used to build up symmetry operators in more general backgrounds. This applies both to cases where we do not necessarily take a large $N$ limit of a given brane probe theory, but also in cases where we engineer a QFT of interest purely in terms of geometry. Our aim in this section will be to take some first steps in this direction, showing how topological symmetry operators constructed from metric isometries fuse and braid with over symmetry operators of a QFT.

As a representative example, we shall be interested in using geometric isometries to engineer examples of duality / triality defects in 4D $\mathcal{N} = 4$ SYM. These sorts of duality / triality defects were constructed in \cite{Kaidi:2021xfk, Choi:2022zal} and given top-down implementations in \cite{Heckman:2022xgu, Bashmakov:2022uek}. In \cite{Heckman:2022xgu}, the duality defect is realized via a IIB 7-brane, and in \cite{Bashmakov:2022uek}, it is constructed in the class $\mathcal{S}$ formalism (see also \cite{DelZotto:2024tae}). The latter geometrizes in IIB, and this is the setup that we will consider more closely. After discussing this specific case, we will state the immediate generalization.

Consider 4D $\mathcal{N}=4$ $\mathfrak{su}(2)$ Super-Yang-Mills as realized by IIB on $\mathbb{R}^{3,1}\times T^2\times \mathbb{C}^2/\mathbb{Z}_2$. The duality defect is constructed from the S-transformation of SL$(2;\mathbb{Z})$ acting on the torus $T^2$ at specific value of complex structure $\tau=i$, and a halfspace gauging. The former can be used to construct a twist defect following \cite{Kaidi:2022cpf}.

\begin{figure}
\centering
\scalebox{0.8}{
\begin{tikzpicture}
	\begin{pgfonlayer}{nodelayer}
		\node [style=none] (0) at (-2, 2) {};
		\node [style=none] (1) at (-2, -2) {};
		\node [style=none] (2) at (2, 2) {};
		\node [style=none] (3) at (2, -2) {};
		\node [style=none] (4) at (2, -2.25) {};
		\node [style=none] (5) at (-2, -2.25) {};
		\node [style=none] (6) at (-2, 2.25) {};
		\node [style=none] (7) at (2, 2.25) {};
		\node [style=Star] (8) at (0, 0) {};
		\node [style=none] (9) at (-2, 0) {};
		\node [style=none] (10) at (-1, -0.75) {};
		\node [style=none] (11) at (-1, -0.125) {};
		\node [style=none] (12) at (-1, 0.125) {};
		\node [style=none] (13) at (-1, 0.875) {};
		\node [style=none] (14) at (-1, -1.125) {$T^2$};
		\node [style=none] (15) at (-0.9, 1.25) {$\text{S}\cdot T^2$};
		\node [style=none] (16) at (-3.25, 0.25) {Physical};
		\node [style=none] (17) at (-3.25, -0.25) {Boundary};
		\node [style=none] (18) at (3.25, 0.25) {Topological};
		\node [style=none] (19) at (3.25, -0.25) {Boundary};
		\node [style=none] (20) at (0.875, 0) {$T^2/\mathbb{Z}_4$};
		\node [style=none] (21) at (0, -3) {};
	\end{pgfonlayer}
	\begin{pgfonlayer}{edgelayer}
		\draw [style=DottedLine] (7.center) to (2.center);
		\draw [style=DottedLine] (4.center) to (3.center);
		\draw [style=DottedLine] (5.center) to (1.center);
		\draw [style=DottedLine] (6.center) to (0.center);
		\draw [style=ThickLine] (2.center) to (3.center);
		\draw [style=ThickLine] (0.center) to (1.center);
		\draw [style=DashedLine] (9.center) to (8);
		\draw [style=ThickLine] (10.center) to (11.center);
		\draw [style=ArrowLineRight] (12.center) to (13.center);
	\end{pgfonlayer}
\end{tikzpicture}
}
\caption{We sketch the geometrization of SymTFT twist defect realizing an S-duality transformation on the torus $T^2$. The defect is defined by an exceptional fiber $T^2/\mathbb{Z}_4$ from which a monodromy branch cut eminates, ultimately terminating on the physical boundary. Crossing the branch cut the $T^2$ is transformed, the monodromy localizes to the branch cut.}
\label{fig:DualityDefect}
\end{figure}
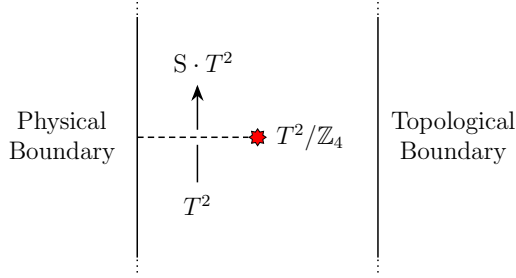

The internal geometry in our setup is $T^2\times \mathbb{C}^2/\mathbb{Z}_2$. In the SymTFT\,/\,SymTh, we now consider a codimension-2 twist defect defined by an S-monodromy. The direct product $\mathbb{R}^{3,1}\times X$ is deformed by a metric defect which, if encircled, maps the torus to the S-transformed torus and otherwise acting trivially on the geometry. See figure \ref{fig:DualityDefect}. In 4D, the S-duality matrix
\begin{equation}
    S=\left( \begin{array}{cc} 0& -1\\ 1 &0 \end{array}\right)\,,
\end{equation}
is order 4 and correspondingly the exceptional fiber is $(T^2/\mathbb{Z}_4)\times \mathbb{C}^2/\mathbb{Z}_2$. With this, we have in the notation of \eqref{eq:collapseeq} that
\begin{equation}
    X=T^2\times (S^3/\mathbb{Z}_2)\,, \qquad  X'=(T^2/\mathbb{Z}_4)\times (S^3/\mathbb{Z}_2)\,.
\end{equation}
Then, deforming both the A-cycle and the B-cycle from $X$ into $X'$, we find that these collapse. This simply follows from $T^2/\mathbb{Z}_4$ topologically being a 2-sphere with three orbifold points (modeled twice on $\mathbb{C}/\mathbb{Z}_2$ and once on $\mathbb{C}/\mathbb{Z}_4$) and consequentially $\text{ker}\,\pi_1\cong \mathbb{Z}^2$.

We now consider a 1-parameter family of A- or B-cycles, which sweep out a non-compact 2-cycle in the geometry with a twist defect inserted. We orient this non-compact 2-cycle as in figure \ref{fig:DualityDefectLike2}. Taking a direct product with a representative for the generator of $H_1(S^3/\mathbb{Z}_2)$, we construct a non-compact 3-cycle. The non-compact direction is parameterized by $x_\perp\geq 0$, which is a spacetime direction normal to the twist defect. This non-compact 3-cycle, which is torsional of order 2, is also ``at infinity" and brane wrappings construct symmetry operators of 4D $\mathcal{N}=4$ $SU(2)$ Super-Yang-Mills theory.

Now, we wrap a D3-brane on this non-compact 3-cycle. This results in a topological 2D (half-)surface in the 4D $\mathcal{N}=4$ $SU(2)$ Super-Yang-Mills theory, parameterized by $x_\perp\geq 0$ and a coordinate $y$ parallel to the duality defect. At $x_\perp=0$ the coordinate $y$ parameterizes a line (topologically $S^1$ or $\mathbb{R}$) which is the terminus of the topological surface on the codimension-1 operator realized by the twist defect. The interacting degrees of freedom of 4D $\mathcal{N}=4$ $SU(2)$ result from D3-branes wrapped on the vanishing 3-cycles which are a product of the A- and B-cycles and the vanishing $\mathbb{P}^1$ of $\mathbb{C}^2/\mathbb{Z}_2$. As such, the initial D3-brane wrapping at infinity is identified as a Gukov-Witten operator (when the B-cycle is used in the construction, when the A-cycle is used the operator is trivialized by the topological boundary conditions).

The consequence of the above, in the 4D spacetime QFT, is that the Gukov-Witten operators of 4D $\mathcal{N}=4$ $SU(2)$ Super-Yang-Mills theory at $\tau=i$ can end on duality defects. This matches the field theory result, Gukov-Witten operators trivialize when crossing the duality defect due to the half-space gauging. In the corresponding half-space, the global form is such that these operators are trivial, and consequently they are seen to terminate on the duality defect.

We also have an immediate generalization. Consider a duality or triality interface in 4D $\mathcal{N}=4$ Super-Yang-Mills theory with simply laced gauge algebra $\mathfrak{g}_{\text{ADE\,}}$ separating two distinct global forms at different $\tau$. Then, the Gukov-Witten operators in either halfspace can terminate on the corresponding twist defect in the SymTFT\,/\,SymTh. Similarly, extensions to more general class S theories are immediate.

We return briefly to the example of 4D $\mathcal{N}=4$ $SU(2)$ Super-Yang-Mills theory. The bottom up construction of the duality defect involves the minimal abelian TFT $\mathcal{A}^{2,1}$ of reference \cite{Hsin:2018vcg}, which exhibits a $\mathbb{Z}_2$ 1-form symmetry acting on lines. Taking the perspective of \cite{Heckman:2022xgu}, we can now identify these lines with respect to the ambient 4D gauge theory. In this reference, the 7-brane used to construct the duality defect is of elliptic type III$^*$ that can be ``higgsed" to $(p,q)$-7-branes of type $A,B,C$. These are characterized similarly geometrically as above and the Gukov-Witten operators can stretch between these, similar to open strings between D-branes. Upon ``unhiggsing"  them back to the duality defect, they result in the lines acted on by the 3D worldvolume 1-form symmetry of the $\mathcal{A}^{2,1}$ theory.

%\newpage

\section{Conclusions \label{sec:CONC}}

One of the general aspects of the AdS/CFT correspondence is that global symmetries of the CFT are dual to gauge symmetries in the bulk AdS theory.
This can be sharpened to the statement that topological symmetry operators of the boundary system are dual to dynamical branes.
In this paper, we have analyzed this statement in the case where the gauge symmetries of the bulk descend from isometries on the internal factor of higher-dimensional spacetimes of the form $\mathrm{AdS} \times X$.
The general procedure we have developed to produce the symmetry operators is to start with a non-topological defect and then boost it so that it is localized in a small sliver close to the boundary.
Detaching from the boundary leads to the presence of some defect-anti-defect fusion products.
In particular, we have shown how to describe the resulting configurations directly via singular fibrations and considered examples of such configurations pushed deeper into the bulk. We have also shown how complementary expectations from top-down / supergravity-based approaches to symmetry theories naturally fit with more ``bottom-up'' considerations based on proposed SymTFT formulations for continuous symmetries.
In particular, we have seen that restrictions in a small sliver of the bulk AdS geometry can produce non-compact gauge groups such as
$\mathbb{R}$. We have given a general prescription for reading off the topological terms of these defects,
and have also taken some preliminary steps in reading off the fusion of these symmetry operators with other topological operators.
In the remainder of this section, we discuss some potential avenues for future investigation.

The general procedure we have outlined works equally well in the case of both abelian and non-abelian symmetry groups.
That being said, it would be exciting to determine further details on the properties of fusion rules in the non-abelian case.
This is especially prevalent in situations with extended supersymmetry.

From the perspective of the 10D / 11D starting point, the isometries constitute a particular class of diffeomorphisms of the higher-dimensional spacetime. Thus, it is natural to ask whether we can use the methods developed here to directly build topological symmetry operators for spacetime symmetries of a QFT. In this vein, it would also be interesting to extend these considerations to discrete symmetries such as parity, time reversal, and more general reflection symmetries (see \cite{Dierigl:2023jdp} for a recent example along these lines).

Much as in \cite{Heckman:2024oot} similar considerations presented here apply to general holographic spacetimes.
In particular, given a continuous symmetry of the non-gravitational dual system, it is natural to use this as a way of inferring properties of a candidate extra-dimensional extension of the bulk gravitational system.

%\newpage

\section*{Acknowledgements}

We thank V. Balasubramanian, V. Chakrabhavi, J. McNamara, B. McPeak, and E. Torres for helpful discussions. JJH thanks the KITP for hospitality during part of this work; this research was supported in part by grant NSF PHY-2309135 to the Kavli Institute for Theoretical Physics (KITP). JJH\ and MH\ thank the 2024 Simons Summer Workshop for hospitality during part of this work. JJH\ thanks the Harvard Swampland Initiative for hospitality during part of this work. JJH and MH thank the Harvard CMSA for hospitality during the completion of this work. MH thanks the UPenn theory group for hospitality during the completion of this work. The work of MC, JJH, and CM is supported by DOE (HEP) Award DE-SC0013528. The work of MC\ and MH\ is supported by the Simons Foundation Collaboration grant \#724069. The work of MC\ is also supported by Slovenian Research Agency (ARRS No. P1-0306) and Fay R. and Eugene L. Langberg Endowed Chair funds. The work of JJH and MH is also supported in part by BSF grant 2022100. The work of MH is supported by the Marie Skłodowska-Curie Actions under the European Union’s Horizon 2020 research and innovation programme, grant agreement number \#101109804. MH acknowledges support from the
the VR Centre for Geometry and Physics (VR grant No. 2022-06593). The work of CM is also supported by the DOE through QuantISED grant DE-SC0020360.

%\newpage

\appendix

\section{Broken Symmetries and Massive Bulk Gauge Fields} \label{app:MASSIVE}

In this Appendix, we briefly discuss some examples of broken symmetry operators for massive bulk gauge fields.
This naturally occurs in the context of the AdS/CFT correspondence since there is typically an entire Kaluza-Klein tower of states after performing reduction on AdS$_{D+1} \times X_{m+1}$.
In particular, one can even carry out consistent truncation schemes where the massless modes and only a few massive modes are retained (see, e.g., \cite{Cvetic:2000dm, Cvetic:2000nc, Gauntlett:2007ma, Gauntlett:2010vu,
Cassani:2010uw, Liu:2010sa, Skenderis:2010vz}).

To begin, we shall assume that we have engineered a CFT via a stack of $N$ coincident branes $\mathcal{B}$ of worldvolume dimension $D$ probing the tip of a cone $Y = \mathrm{Cone}(X)$.
In the large $N$ holographic dual, this results in a geometry of the form AdS$_{D+1} \times X_{m+1}$. We assume
that $X_{m+1}$ has a continuous isometry, and an associated $S^1$ fibration.

Next, we construct a class of heavy defects corresponding to codimension-2 defects in the boundary CFT.
A natural way to do this is to introduce another brane of the same type used to build the CFT$_{D}$ in the first place;
we can have it fill the subspace AdS$_{D-1} \times S^{1}$ inside of AdS$_{D+1} \times X_{m+1}$.
Observe that this configuration is stable due to the dynamics of the brane.
In fact, this sort of heavy defect was used in \cite{Gukov:2006jk} to engineer examples of Gukov-Witten operators in $\mathcal{N}=4$ Super Yang-Mills theory. Recall that these surface operators are labeled by elements of the corresponding Lie algebra.
This is encoded in the heavy defect through a choice of background gauge fields and worldvolume scalar fields.

In the bulk, there is a natural object that links with this defect.
In the present background, we can consider the magnetic dual brane $\widetilde{\mathcal{B}}$
wrapped on a subspace $M_{m-1} \subset X_{m+1}$ which links with the distinguished $S^{1}$.
In the AdS$_{D+1}$ directions this specifies the worldline of a heavy particle.
Of course, this is nothing but
the ``Giant gravitons'' of references
\cite{McGreevy:2000cw, Balasubramanian:2001nh, Hashimoto:2000zp}.
They do not collapse because they have non-trivial angular momentum on $M_{m-1}\subset
X_{m+1}$.
As explained in reference \cite{Mikhailov:2000ya}, one way to figure out the orbit of these objects is to start with a supersymmetric cycle in $Y = \mathrm{Cone}(X_{m+1})$.
This cuts out a subspace $M_{m-1}\subset X_{m+1}$.
The time evolution is then obtained by taking the corresponding isometry and evolving with respect to it, producing a family of orbiting solutions which we label as $M_{m-1}(\lambda)=J_{\lambda}(M_{m-1})$,
i.e., we consider the finite time evolution as generated by the rotation $J_{\lambda}$.

Given this set of objects, it is natural to ask what happens when we apply the same ``boosted defects'' procedure introduced in section \ref{sec:BOOST}. Consider first the heavy defect.
After boosting, this indeed produces a codimension-1 object in the boundary system.
Likewise, the orbiting giant graviton can instead be replaced by a heavy line operator in the bulk which terminates on a local operator in the boundary system.
One might therefore be tempted to identify the codimension-1 object with a symmetry operator, and the heavy line operator in the bulk with its linking counterpart in the
bulk SymTFT (see reference \cite{Waddleton:2024iiv}).

However, there is an important subtlety with this proposal.
The issue is that we have actually constructed a candidate symmetry operator for a broken symmetry!
Said differently, this symmetry is not really present in the dual CFT.
One can see the issue in a few different ways.
One way to observe a potential issue is to observe that in a consistent truncation scheme on $X_{m+1}$, $M_{m-1}$ does not define a stable cycle in $H_{\ast}(X_{m+1}, \mathbb{Z})$.
In particular, a reduction of an $m$-form potential over $M_{m-1}$ can at best produce a massive gauge field in the bulk.
This can be corroborated by a careful analysis of consistent truncation, where one finds that one keeps the KK gauge boson $A_{1}$ as well as a massive counterpart $\mathbb{A}_{1}$.
The actual massless gauge field turns out to be a linear combination of $A_{1}$ and $\mathbb{A}_{1}$ (see in particular reference \cite{Liu:2010sa} as well as \cite{Gauntlett:2010vu, Cassani:2010uw}).

One can also directly see this candidate $U(1)$ being broken in the bulk by using our previously constructed giant graviton state.
Indeed, the radially extending heavy line sits at some fixed time slice in the AdS$_{D+1}$.
We can consider the worldline of a giant graviton which terminates at some finite radial profile.
This is a clear indication that in the bulk, the candidate line operator is not a genuine defect.

Finally, let us turn to the interpretation in the dual CFT.
In the bulk, we have a massive gauge field $\mathbb{A}_{1}$. In the boundary theory, we should thus expect a spin-1 operator, but with a scaling dimension $\Delta > \Delta_{\text{current}}$, where $\Delta_{\text{current}} = (D-1)$ is the scaling dimension for a conserved current.
These are of course interesting spin-1 states to consider, but the increase in dimension signals that they are at best associated with a broken gauge symmetry in the bulk.

%\newpage

\bibliographystyle{utphys}
\bibliography{NoGloboII}

\end{document}